\pdfoutput=1 %
\documentclass[a4paper,twoside,11pt]{article}

\usepackage[USenglish]{babel} %
\usepackage[T1]{fontenc}

\usepackage{graphicx}
\usepackage{url}
\usepackage[labelsep=endash, figurename=Figure, tablename=Table, textfont=it]{caption}
\usepackage{subcaption}
\usepackage{float}  
\usepackage{amsmath}
\usepackage{hyperref}

\usepackage{xcolor}

\usepackage[top=2.5cm, bottom=2.5cm, left=2.5cm, right=2.5cm]{geometry}
\usepackage[authoryear,round]{natbib}
\usepackage{placeins}
\usepackage{titling}
\usepackage{booktabs}
\usepackage{longtable}
\usepackage{makecell}
\usepackage{siunitx}
\usepackage{authblk}

\definecolor{darkblue}{rgb}{0,0,0.5}
\hypersetup{colorlinks=true,linkcolor=black,citecolor=darkblue,urlcolor=darkblue}

\usepackage{tcolorbox}
\tcbuselibrary{breakable,skins}
\usepackage{rotating}

\usepackage{xurl} %
\usepackage{enumitem}
\usepackage{array}
\newtcolorbox{rulebox}[1][]{enhanced, colback=white, colframe=black!70,
  fonttitle=\bfseries\small, title={#1}, left=4pt, right=4pt, top=3pt, bottom=3pt, fontupper=\small}
\newtcolorbox{ruleboxlong}[1][]{breakable, enhanced, colback=white, colframe=black!70,
  fonttitle=\bfseries\small, title={#1}, left=4pt, right=4pt, top=3pt, bottom=3pt, fontupper=\small}
\newtcolorbox{promptbox}[1][]{breakable, enhanced, colback=white, colframe=black!70,
  fonttitle=\bfseries\small, title={#1}, left=4pt, right=4pt, top=3pt, bottom=3pt,
  fontupper=\ttfamily\scriptsize, fontlower=\ttfamily\scriptsize,
  halign=flush left, halign lower=flush left, segmentation style={dashed}}
\newcommand{\pnote}[1]{{\normalfont\itshape(#1)}}
\newcolumntype{L}[1]{>{\raggedright\arraybackslash}p{#1}}
\title{When Disclosure Is Compulsory: Evaluating Ride-Hailing Rebalancing Policies with Self-Interested LLM Driver Agents}
\author[1,*]{Xinyu Wang}
\author[1]{Andres Fielbaum}
\affil[1]{School of Civil Engineering, The University of Sydney, Darlington 2008 NSW, Australia}
\affil[*]{Corresponding author: xinyu.wang@sydney.edu.au}
\date{}

\begin{document}
\maketitle

\begin{abstract}
The EU Platform Work Directive, which member states must transpose by December 2026, obliges ride-hailing platforms to state in plain language the main parameters of their automated decisions. Since plain language is exactly what a large language model (LLM) can read, and LLMs are increasingly used as personal decision-making assistants, we ask how rebalancing policies perform once their rules are disclosed to drivers who use AI to make decisions.

We compare two mainstream rebalancing methods: a reactive policy (R), which sends idle drivers to cover the requests just lost, and an anticipatory policy (A), which plans moves on a demand forecast. Each driver is modeled as an LLM agent that maximizes its own net income and receives either no rules or the rules in the detail the directive requires. We tested on two models with different scales: a three-zone model with ten drivers which can be solved to optimality, and Manhattan network with 4091 nodes, 29 hexagonal zones and 500 drivers, which is closer to a realistic city. Once the rules are disclosed, the drivers follow the platform more closely: in Manhattan, the share of rebalancing offers declined falls from about 80\% to 25\% under both policies. To the platform, disclosure costs R 3.7\% of its profit, while A's profit is unchanged. Against full compliance, the self-interested drivers make A a `win-win-win': driver income, platform profit and the service rate are all raised by about 10--12\% each. Under mixed scenarios when only some drivers use an AI agent and others follow the instruction of the platform, the game among drivers is close to zero-sum under R, where AI drivers take fares from compliant drivers, and positive-sum under A, where they serve additional passengers, who would otherwise be lost.

Under mandatory disclosure, rebalancing policies that behave equally on profit when drivers comply can show opposite results, and the policy decides whether the gain of AI drivers is the loss of compliant drivers, or a `win-win-win' can occur. To the best of our knowledge, this is the first study to use LLM agents to evaluate how disclosure affects rebalancing policies in ride-hailing.
\end{abstract}


\section{Introduction}
\label{sec:intro}

Ride-hailing platforms allocate work, set prices and move drivers by algorithm; however, the mechanism behind them has not been fully explained to the workers. Directive (EU) 2024/2831 on platform work changes this. Its Article 9 requires a platform to inform workers of the main parameters of its automated decision-making systems, the relative importance of those parameters, and the way in which a worker's behaviour affects the decisions; Article 2 places pay and the pricing of tasks within scope \citep{eu2024directive,adamsprassl2023blueprint,veale2023fortifying,rosin2025transparency}. The obligation arrives in a workforce that already tries to work the rules out for itself. Studies of algorithmic management show Uber drivers inferring matching and control rules they cannot see \citep{mohlmann2021algorithmic}, assembling a shared reading of the algorithm from clues and tests \citep{mohlmann2023sensemaking}, and asking most often for the rules on pricing, promotions, routing and dispatch \citep{rao2025rideshare}. Thus, the question for a platform operator is no longer whether to disclose, but what happens once it must.

Drivers who understand a rule act on it, and the dispatch rule decides what their reaction does to everyone else. Drivers have logged off together to trigger surge pricing \citep{schroder2020anomalous,yu2023cycles,bai2023hiding,tripathy2023collusion}, cherry-pick requests when the surge mechanism rewards it \citep{garg2022surge,ashkrof2022acceptance}, and reposition strategically across a network \citep{afeche2023strategic,wang2025relocation}. 
Rebalancing, the platform's own instrument for moving idle drivers, has been studied from the opposite side. Reactive policies use only the current state, covering the requests that were just lost \citep{alonsomora2017ondemand} or applying backpressure to the current imbalance \citep{kanoria2024mirror}, where anticipatory policies plan moves on a demand forecast \citep{zhang2016mpc,iglesias2018datadriven,engelhardt2023predictive}. Almost all of this work assumes that drivers comply with the moves they are sent, and the few studies that relax the assumption estimate or adapt to driver behaviour rather than disclose the rule \citep{chen2024irebalance,brar2024adherence}. How much drivers will understand the disclosed rule, and how they will act on it, is therefore unknown. 

As required by law, the disclosed rule will be plain text, which is a natural input for a large language model (LLM). An LLM agent can take the rule as written, weigh it against the market state it is shown and choose an action, without a specified reward function or retraining as Reinforcement Learning (RL) requires. With the rapid development of LLMs, they are already being used as economic agents in laboratory settings \citep{horton2023homo,chen2023rationality}, automated decision-makers in markets \citep{xu2026llmagents}, and delegated decision-makers for people \citep{anthropic2026projectdeal}. It is therefore plausible that, once the platform discloses its rules, human drivers can use an LLM agent to make decisions in their own interest. Thus, to evaluate the behaviour of self-interested drivers under such disclosure, this study does not need to ask the LLM to imitate a human driver, but to act as a proxy that a driver could use, whose only objective is the driver's own net income.

The paper makes three contributions. First, to the best of our knowledge, it is the first study of how drivers act, through an LLM agent, on disclosed rules described in plain language. Second, it shows that stable and general patterns exist across different LLMs and different policies, which reveal intrinsic properties of the on-demand mobility system. Third, it gives operators guidance for the soon-to-be-compulsory disclosure of policies at a time when LLMs are widely used: operators should reconsider and redesign their policies under information equity, and this study provides a framework to evaluate the performance of different policies under the new information disclosure requirement, from a small laboratory model to a realistic city-scale simulation.

The rest of the paper is organised as follows. Section~\ref{sec:literatures} reviews the literature in five parts, from the law that requires disclosure to LLM agents in markets, and states the research gaps that the contributions address. Section~\ref{sec:method} describes the ride-hailing system and the two rebalancing policies (Section~\ref{sec:market}), and the AI driver agents, including how the policies are disclosed to them (Section~\ref{sec:llm}). Section~\ref{sec:setup} sets up the small model, the city model, the comprehension test and the metrics of evaluation, and Section~\ref{sec:results} reports the results: the comprehension test and the small model, in which the two LLMs are compared, then the city under the reactive and the anticipatory policy, a comparison of the two policies, and city markets in which only some drivers use an AI agent. Section~\ref{sec:discussion} summarises the findings, draws the consequences for operators and regulators, and points to future work. The appendices give the parameter values, the prompts and disclosed rules, the models and their costs, the comprehension questions, and further results for the small model and the city.

\section{Literature Review}
\label{sec:literatures}

This section reviews the literature from five perspectives, each of which studies one necessary part of this study. It starts with the law which forces information disclosure (Section~\ref{sec:rw-law}), then turns to the strategic behaviour of drivers who know the rules (Section~\ref{sec:rw-strategic}), the literature on rebalancing methods (Section~\ref{sec:rw-rebalancing}), the information design literature, which shows that the design of a mechanism affects the results (Section~\ref{sec:rw-disclosure}), and finally the LLM agent literature (Section~\ref{sec:rw-llm}). Lastly, Section~\ref{sec:rw-gap} states the research gaps and the contributions of this study that address them.

\subsection{Rules must be disclosed: algorithmic management and the Platform Work Directive}
\label{sec:rw-law}

Two strands explain why ``what happens after disclosure'' has only now become a question. The first is legal. \citet{adamsprassl2023blueprint} set out a blueprint for regulating algorithmic management in which transparency, explanation and human review are the pillars; this is the thinking behind Article 9 of Directive (EU) 2024/2831, which turned the main parameters of automated decisions, their relative importance and the effect of a worker's behaviour on them into an obligation. \citet{veale2023fortifying} examine how ``main parameters'' should be read and enforced, and \citet{rosin2025transparency} compare the directive with the GDPR, the Platform-to-Business Regulation and the AI Act: the directive widens transparency rights beyond all three, but asks for parameters and logic, not formulas. Our disclosure condition L is written to that level. The second strand is organisational and human--computer interaction research on how drivers live with rules they cannot see. \citet{mohlmann2021algorithmic} describe Uber drivers under algorithmic matching and control working out rules from their consequences; \citet{mohlmann2023sensemaking} theorise this as algorithm sensemaking, a multi-step process of enacting, selecting and retaining interpretations of the algorithm; and \citet{rao2025rideshare}, analysing over a million posts in driver forums, find that pricing, promotions, routing and dispatch rules are what drivers most want disclosed. Our condition N is the situation these studies describe, state visible and rules hidden, and L replaces the guessing with the text the law requires.

Both strands judge transparency as a form of worker protection. Neither asks how a self-interested driver would use a disclosed rule, and neither compares dispatch rules once they are disclosed. To evaluate such influence, we put the directive-level text into the experiment as written.

\subsection{Driver knows the rule: Strategic behaviour and manipulable pricing}
\label{sec:rw-strategic}

The nearest neighbours of this paper study drivers who understand surge pricing. \citet{yu2023cycles} show that when prices track online supply, drivers who log off at low prices and return after the surge can form a Nash equilibrium that produces price cycles, and that a price floor removes them. \citet{bai2023hiding} model providers who hide supply to trigger surge and give, in their Lemma~1, the condition under which collusion does not pay; \citet{tripathy2023collusion} characterise coordinated log-offs under a known surge rule and the bonus or freeze-period remedies. In each case the known rule is the precondition for the manipulation; we turn that precondition into a treatment. \citet{garg2022surge} show that multiplicative surge is not incentive compatible for drivers, who cherry-pick trips, and derive an incentive-compatible driver surge; \citet{ma2022stp} give a public spatio-temporal pricing mechanism under which accepting every dispatch is a subgame-perfect equilibrium, the natural reference for a future incentive-compatible anticipatory design. On the empirical side, \citet{ashkrof2022acceptance} estimate from stated-preference data how pickup time, surge income and other attributes govern ride acceptance, and \citet{schroder2020anomalous} document anomalous supply shortages induced by dynamic pricing at 137 locations worldwide.

The theoretical studies take the rule as known and assume that drivers respond to it with perfect rationality, and the empirical ones observe drivers without varying what they are told about the rule. Disclosure is therefore never a variable, and how a driver who is not perfectly rational responds once a rule is disclosed remains unknown. 

\subsection{Rebalancing: Reactive and Anticipatory}
\label{sec:rw-rebalancing}

There are two mainstream families of rebalancing methods in ride-hailing, and this study evaluates both of them. Reactive rebalancing, which sends idle vehicles to the requests that were just lost at minimum travel, is the rebalancing step of \citet{alonsomora2017ondemand}; \citet{kanoria2024mirror} show that mirror backpressure, a policy that uses only the current state and no demand rates, is near-optimal in steady state. Anticipatory rebalancing, which optimizes vehicle movements over a horizon on a demand forecast, is the model predictive control (MPC) of \citet{zhang2016mpc} and its data-driven form in \citet{iglesias2018datadriven}. The survey of \citet{zardini2022amod} organises autonomous mobility-on-demand (AMoD) control into steady-state flow, MPC and learning-based methods, and our two policy classes follow that taxonomy. Anticipatory repositioning usually wins such comparisons, as in the ride-pooling simulation of \citet{engelhardt2023predictive}, which also serves as the implementation reference for our city model. Anticipation is not limited to the moves of idle vehicles: \citet{wang2026rlrouting} build it into routing and matching and learn its weight with reinforcement learning, and \citet{wang2026walking} use short-horizon demand to choose where the last passenger is dropped off. 

A smaller literature lets drivers refuse or act strategically. \citet{afeche2023strategic} analyse a network fluid model with self-interested drivers and show how admission and repositioning control shape the equilibrium; \citet{wang2025relocation} study strategic relocation in a leader--follower game and identify when relocation subsidies are redundant because drivers already move where the platform wants them, which echoes our finding that accurate repositioning is a win-win. \citet{chen2024irebalance} drop the obedience assumption altogether and estimate acceptance of repositioning recommendations from a field study with 99 drivers; \citet{brar2024adherence} plan rebalancing under adherence that evolves with the drivers' experience. The conflict between the operator's objective and that of individual participants is not limited to drivers: \citet{wang2027meetingpoint} show that operator-optimal and user-optimal meeting points in ridepooling diverge systematically, and that the cost of the disagreement concentrates in the zones where the two disagree most and falls more on users than on the operator.

AMoD repositioning assumes vehicles obey. The work on strategic or uncertain compliance treats driver behaviour as something to estimate and adapt to; none of it makes disclosure a variable or compares reactive and anticipatory rules after disclosure. We make declining an offer a driver action, report who pays for it, and claim that the same driver behaviour has opposite effects on others under the two rules.

\subsection{Influence of disclosure: strategic-response theory and information design}
\label{sec:rw-disclosure}

Theory explains why the answer should depend on the rule rather than on disclosure itself. In strategic classification and mechanism design, the decision rule is told to those it judges. \citet{wang2023transparency} show that with strategic users, disclosing the algorithm often benefits the firm but not necessarily the users. \citet{ghalme2021dark} define the price of opacity when agents must act on a guessed classifier, and \citet{shao2025reveal} show that hiding the classifier can backfire in the online setting. \citet{gonczarowski2023descriptions} show that the same strategy-proof mechanism, described in different ways, changes how people play. Information design in ride-hailing asks how much the platform should tell drivers about demand: \citet{candogan2023information} derive the optimal disclosure of the demand state to steer repositioning, and \citet{lian2026transparency} argue that transparency, control and pay must be judged jointly, the same direction as our finding that the effect of disclosure depends on dispatch.

In this theory the rule is a classifier or a static mechanism with no dynamic, spatial or operational outcome. And none of the studies treats the rule as a text that an agent can read and act on. 

\subsection{AI agents: LLMs in markets and mechanisms}
\label{sec:rw-llm}

\citet{horton2023homo} propose LLMs as simulated economic agents that can be given endowments, information and preferences. \citet{chen2023rationality} find that GPT passes revealed-preference rationality tests at a higher level than human subjects. \citet{akata2025repeated} find that LLMs do well in self-interested repeated games and badly in games that require coordination, a reason to expect individual, uncoordinated self-interest from our agents. \citet{fish2026collusion} show that GPT-4 pricing agents reach supra-competitive prices when they keep notes across periods; our agents have no memory, so our result is a lower bound on learned collusion. \citet{kleinberg2021monoculture} show that many decision-makers sharing one algorithm make correlated errors, the theory behind the herding risk of a fleet driven by one model. \citet{hoshino2026matching} find that LLM agents in strategy-proof matching mechanisms mostly report truthfully and deviate more as the rules become complex. In transport, \citet{liu2025llmabm} propose a framework for modelling transport systems with LLM agents as travellers, and a line of work aligns such agents with observed behaviour: \citet{liu2026persona} align them with mode-choice data through learned traveller personas, \citet{liu2026dualagent} use a second LLM agent to calibrate traveller agents to day-to-day route choice, and \citet{yan2026vot} find human-like sensitivities in the value of travel time of LLMs but differences across models in their magnitude. In platform labour, \citet{xu2026llmagents} survey LLM agents as behavioural simulators and decision-support tools for transportation-enabled service platforms, and \citet{zhang2026llmdr} simulate on-demand delivery with persona-grounded LLM riders. All of these aim at realism rather than at rule exploitation. Outside the academic literature, \citet{anthropic2026projectdeal} report an experiment in which people delegated trades to AI agents and stronger models systematically won.

No study has had AI driver agents read the rules a platform actually runs and checked whether they exploit them; the LLM collusion and mechanism studies are abstract markets without spatial, dynamic outcomes; and platform-side LLM dispatch work treats the model as a tool, not as a self-interested driver. Our agents read a rule a platform actually runs and act on it in a dynamic, spatial market.

\subsection{Research gaps and contributions}
\label{sec:rw-gap}

To the best of our knowledge, the literature reviewed above leaves three gaps. Each is stated below together with how this paper addresses it.
\begin{enumerate}
\item \textbf{Disclosure in plain language to self-interested agents.} Information equity between the platform and its drivers has not been studied in the form the directive requires: the rule written in plain language and read by a self-interested agent who acts on it. The studies of strategic drivers and strategic response either take the rule as known to rational agents or disclose a classifier, the demand state or a forecast (Sections~\ref{sec:rw-strategic} and~\ref{sec:rw-disclosure}), and the LLM-agent studies aim at realism rather than at acting on a disclosed rule (Section~\ref{sec:rw-llm}). We address this gap by disclosing the rules in plain language, at the directive's own level, to LLM driver agents whose only objective is their own net income, and by measuring how the agents act on them.
\item \textbf{Generality across models, policies and scales.} Whether the response to a disclosed rule is a property of the system or of one model and one setting is unknown. LLMs differ from one another in the size of their economic trade-offs \citep{yan2026vot}, and the theory of strategic drivers and strategic response is built on stylised models (Sections~\ref{sec:rw-strategic} and~\ref{sec:rw-disclosure}). We address this gap by running two unrelated LLMs through the comprehension test and the small model under both policies, where they understand the rules equally well and respond to disclosure in the same direction.
\item \textbf{Policies under the same disclosure.} How different policies perform under the same disclosure has not been studied: rebalancing policies are compared with drivers who comply, and the few studies that relax compliance do not disclose the rule (Section~\ref{sec:rw-rebalancing}), so whether a policy that works under compliance still works once its rule is disclosed is unknown. We address this gap by comparing reactive and anticipatory rebalancing under the same disclosure, which gives operators a framework, from a small laboratory model to a city-scale simulation, to evaluate and redesign their policies before disclosure becomes compulsory.
\end{enumerate}

\section{Methodology}
\label{sec:method}

This section describes the model in two parts. Section~\ref{sec:market} sets out the ride-hailing system (Section~\ref{sec:system}) and the two rebalancing policies compared, reactive (R) and anticipatory (A) (Section~\ref{sec:policies}). Section~\ref{sec:llm} describes the drivers as AI agents: what they observe (Section~\ref{sec:inputs}), how the platform's rules are disclosed to them (Section~\ref{sec:disclosure}), which decisions they take (Section~\ref{sec:decisions}), and the two language models that take them (Section~\ref{sec:models}). The two policies differ in one module of the platform and in one paragraph of the disclosed rules; fares, commission, compensation and every other part of the prompt are shared, so a difference between R and A can only come from dispatch. The two simulation setups on which the method is run, their parameters and the benchmarks are given in Section~\ref{sec:setup}. Table~\ref{tab:notation} summarises the notation.

\begin{table}[htbp]
\centering
\footnotesize
\caption{Notations used in this study. Parameter values are given in Table~\ref{tab:params} (Appendix~\ref{app:params}).}
\label{tab:notation}
\begin{tabular}{@{}L{1.9cm} L{5.3cm} L{1.9cm} L{5.3cm}@{}}
\toprule
Symbol & Meaning & Symbol & Meaning \\
\midrule
\multicolumn{2}{@{}l}{\textbf{Indices, policies and conditions}} & \multicolumn{2}{l}{\textbf{Requests and demand}} \\[1pt]
$o$, $d$, $z$ & Zones: origin, destination, any zone & $q_z$ & Requests in zone $z$ over the last price interval \\
$r$; $o_r$, $d_r$ & Passenger request; its origin and destination zones & $q^{\text{lost}}_d$ & Requests lost in zone $d$ since the last plan \\
$v$; $V$ & Vehicle and its driver; set of vehicles & $\hat q_{odk}$ & Forecast requests from $o$ to $d$ in interval $k$ \\
$t$ & Decision step (a period in the small model) & $\lambda(t)$, $\lambda_{od}(t)$ & Mean requests per period, in total and from $o$ to $d$ (small model) \\
$k$ & Interval of A's horizon, $k=0,\dots,H-1$ & $\lambda_0$, $\lambda_1$ & Base level and peak height of $\lambda(t)$ \\
R, A & Reactive and anticipatory rebalancing & $t^{\text{am}}$, $t^{\text{pm}}$, $\sigma$ & Morning and evening peak periods; peak width \\
N, L & No disclosure; law-level disclosure & $w_{od}(t)$ & Share of the period's requests going from $o$ to $d$ \\
 & & $\rho$ & Share of the taxi records sampled as requests (city) \\
\addlinespace[5pt]
\multicolumn{2}{@{}l}{\textbf{Time and distance}} & \multicolumn{2}{l}{\textbf{Fleet and drivers}} \\[1pt]
$\Delta T$ & Length of a decision step (matching) & $f_V$ & Fleet size \\
$\Delta T_p$, $\Delta T_r$, $\Delta T_i$ & Length of a price interval; of a planning interval; of the idle-decision interval & $c$ & Vehicle capacity \\
$H$ & Horizon of A, in price intervals & $s_z$ & Drivers now available in zone $z$ \\
$t_{od}$ & Travel time between the centres of $o$ and $d$; $t_{oo}=0$ & $\bar s_z$ & Idle drivers counted in zone $z$ over the last price interval \\
$\tilde t_{od}$ & $t_{od}$ counted in price intervals, at least one & $s^{\text{arr}}_z$ & Drivers committed to arrive in zone $z$ before the next price update \\
$t_w^{\max}$ & Maximum pick-up waiting time & $s_{zk}$, $s^{\text{arr}}_{zk}$ & Available drivers and committed arrivals in zone $z$, interval $k$ of A's plan \\
$t_{w,r}$ & Waiting time of request $r$, from the request to the pickup & $I_v$ & Net income of driver $v$ over the simulated day \\
$\ell_{od}$ & Distance of a trip from $o$ to $d$ & $\mathcal{R}_v$ & Requests served by driver $v$ \\
$\ell_{vd}$ & Distance from vehicle $v$'s position to the centre of zone $d$; $\ell_{od}$ for a driver in zone $o$ (small model) & $\mathcal{O}_v$; $d_j$ & Rebalancing offers $j$ that driver $v$ accepts and completes; destination zone of offer $j$ \\
$\ell_v$, $T_v$ & Distance driven and time online of driver $v$ & & \\
\addlinespace[5pt]
\multicolumn{2}{@{}l}{\textbf{Prices and payments}} & \multicolumn{2}{l}{\textbf{Costs}} \\[1pt]
$g_z$ & Shortage of zone $z$, between 0 and 1 & $\beta_\ell$, $\beta_T$ & Driver cost per unit of distance driven; per unit of time online \\
$p$ & Base fare rate per unit of distance & $C_v$ & Cost of driver $v$ \\
$\text{fare}_{od}$ & Fare of a trip from $o$ to $d$ & $\beta_w$ & Value of waiting time, per unit of time \\
$\overline{\text{fare}}_{od}$ & Fare of an average trip from $o$ to $d$ at $o$'s current rate & $C_{u,r}$ & Served user cost of request $r$ (Eq.~\ref{eq:usercost}) \\
$\theta$; $\Pi$; $\Delta\Pi$ & Platform commission, as a share of the fare; platform profit over the simulated day; its change against everyone complying & & \\
$b_{vd}$; $\bar b_{od}$ & Compensation of an offer to driver $v$ for zone $d$ (Eq.~\ref{eq:compgen}); of a move between the centres of $o$ and $d$, used in A's plan & \multicolumn{2}{l}{\textbf{Decision variables of the rebalancing plans}} \\[1pt]
$b_0$, $\kappa_{vd}$ & Base compensation per unit of distance; surge compensation per unit of the destination's shortage & $x_{od}$, $x_{odk}$ & Rebalancing moves from $o$ to $d$: under R; under A in interval $k$ \\
$\alpha$ & Compensation as a share of the destination's fare rate per km (city) & $y_{odk}$ & Passenger trips from $o$ to $d$ served in interval $k$ (A) \\
\bottomrule
\end{tabular}
\end{table}

\subsection{Ride-hailing and rebalancing}
\label{sec:market}

In ride-hailing, a platform matches passenger requests to drivers, sets the fare of each trip and keeps a commission; the drivers are independent, keep the rest of the fare and pay the cost of driving. Requests and drivers are rarely in the same places at the same time. In the morning, for example, requests start in residential zones while drivers who have just dropped passengers off collect in the business district, and in the evening the flow is reversed. Rebalancing is the platform's tool for this imbalance: it moves idle drivers from zones with more drivers than requests to zones where requests are going unserved. Here it does so by sending rebalancing offers, which pay a driver a compensation for driving to another zone. Section~\ref{sec:system} describes the system the two policies share, and Section~\ref{sec:policies} the two ways of deciding where the offers go.

\subsubsection{The ride-hailing system}
\label{sec:system}

\paragraph*{Zones and time.} The service area is divided into zones, indexed $o$, $d$ or $z$ (origin, destination, any zone). A trip or move from zone $o$ to zone $d$ takes the travel time $t_{od}$ between the zone centres, with $t_{oo}=0$. Time advances in decision steps $t$ of length $\Delta T$, and the platform works on three clocks: at every decision step it matches waiting requests to idle drivers; at every price interval, of length $\Delta T_p$, it updates each zone's fare and the compensation of offers into it; and at every planning interval, of length $\Delta T_r$, it runs the rebalancing policy and sends rebalancing offers. 

\paragraph*{Requests and matching.} A passenger requests a trip $r$ from an origin zone $o_r$ to a destination zone $d_r$ without rejecting any service allocated by the platform. The fleet is a set $V$ of $f_V$ vehicles $v$, and each carries one request at a time. A waiting request is assigned to an idle driver who can reach the passenger within the maximum pick-up waiting time $t_w^{\max}$, and then the driver can accept or reject it. A rejected request goes to another eligible driver, and the service will be rejected after a fixed number of rejections or when no eligible driver is left.

\paragraph*{Fare.} The fare for each trip composes a base rate which is proportional to the trip length, and a surge term that rises with the shortage of drivers in the origin zone.
\begin{equation}
g_z=\frac{\max\!\big(0,\;q_z-\bar s_z-s^{\text{arr}}_z\big)}{q_z}\qquad(g_z=0\ \text{if}\ q_z=0),
\label{eq:shortage}
\end{equation}
where $q_z$ is the number of requests in zone $z$ over the last price interval $\Delta T_p$, $\bar s_z$ the idle drivers counted in the zone over that interval and $s^{\text{arr}}_z$ the drivers committed to arrive before the next update, either finishing a passenger trip there or driving there on an accepted rebalancing offer. The fare of a trip from $o$ to $d$ is
\begin{equation}
\text{fare}_{od}=p\,(1+g_o)\,\ell_{od},
\label{eq:fare}
\end{equation}
with $p$ the base rate per unit of distance and $\ell_{od}$ the distance of the trip, so the fare lies between $p\,\ell_{od}$ and $2p\,\ell_{od}$. The fare is fixed when the request is made and paid when the trip is completed. The platform keeps a commission $\theta$ of the fare as income.

\paragraph*{Driver cost.} The cost of a driver comes from two sources: the distance it drives and the time it is online, whether carrying a passenger, moving to a zone or moving on its own. The cost is linear in both, with coefficients $\beta_\ell$ and $\beta_T$.
\begin{equation}
C_v=\beta_\ell\,\ell_v+\beta_T\,T_v,
\label{eq:cost}
\end{equation}
where $\ell_v$ is the distance driver $v$ drives for any reason (passenger trip, pickup, rebalancing move or own move) and $T_v$ its time online. This cost enters the driver's net income, Eq.~\eqref{eq:income}.

\paragraph*{Driver actions.} A driver can accept or reject a passenger request, accept or decline a rebalancing offer, and stay or move to another zone on their own. There is no offline option. A driver moving on their own earns no compensation.

\paragraph*{Rebalancing offers and compensation.} The rebalancing policy (Section~\ref{sec:policies}) decides which driver should at which time move to which zone. Each such move becomes a rebalancing offer to one idle driver at a time; a declined offer goes to another idle driver in the same zone. An offer to driver $v$ for zone $d$ pays the compensation
\begin{equation}
b_{vd}=b_0\,\ell_{vd}+\kappa_{vd}\,g_d,
\label{eq:compgen}
\end{equation}
a base payment $b_0$ per unit of the distance $\ell_{vd}$ from the driver's position to the centre of zone $d$, plus a surge payment that rises with the shortage $g_d$ of the destination at rate $\kappa_{vd}$. The coefficients are set so that $\beta_\ell\,\ell_{vd}<b_{vd}<(1-\theta)\,p\,\ell_{vd}$: an offer always covers the driving cost and never pays more than the driver's share of a fare over the same distance, so a move is worth making but never worth more than a passenger. Accepting an offer locks its compensation, which is paid on arrival; only the holder of an offer receives it, and there is no daily cap. Fares, commission, costs and the compensation of Eq.~\eqref{eq:compgen} are identical under the two policies. What the two policies decide differently is which offers to send: how many drivers to move, from which zone to which, and at what moment, which Section~\ref{sec:policies} explains.

\subsubsection{Rebalancing}
\label{sec:policies}

Both policies share the same decision frequency, information about the system, and pricing and offer mechanics, but differ in what the plan responds to and what it optimizes. The reactive (R) method responds to the requests that have just been lost and covers them with the least travel; the anticipatory (A) method responds to the requests it forecasts and maximizes the platform's expected profit. The two follow the two mainstream families in the literature (Section~\ref{sec:rw-rebalancing}). 

\paragraph{Reactive rebalancing (R).}
\label{sec:R}

Following the rebalancing step of \citet{alonsomora2017ondemand}, each plan counts the requests lost in each zone since the last plan and sends idle drivers to cover them with the least travel:

\begin{equation}
\min_{x\ge 0}\ \sum_{o,d} t_{od}\,x_{od}
\ \ \text{s.t.}\ \ 
\sum_{d} x_{od}\le s_o,\quad
\sum_{o} x_{od}\le q^{\text{lost}}_d,\quad
\sum_{o,d} x_{od}=\min\Big(\sum_d q^{\text{lost}}_d,\ \sum_o s_o\Big),
\label{eq:R}
\end{equation}

where $q^{\text{lost}}_d$ is the number of requests lost in zone $d$ since the last plan; $s_o$ is  the drivers now available in zone $o$; and $t_{od}$ is the travel time between zone centres. Each off-diagonal $x_{od}$ becomes rebalancing offers to drivers in zone $o$, one per lost request.

\paragraph{Anticipatory rebalancing (A).}
\label{sec:A}

Following the zonal model predictive control of \citet{zhang2016mpc} and \citet{iglesias2018datadriven}, each plan optimizes driver movements over the next $H$ price intervals to maximize expected net income of the platform, which is constituted by the expected commission minus compensation paid to drivers,
\begin{equation}
\begin{aligned}
\max_{x,y,s\ge 0}\;& \sum_{k=0}^{H-1}\sum_{o,d}\Big(\theta\,\overline{\text{fare}}_{od}\,y_{odk}-\bar b_{od}\,x_{odk}\Big)\\
\text{s.t.}\;& y_{odk}\le \hat q_{odk},\qquad \sum_d\big(x_{odk}+y_{odk}\big)\le s_{ok},\\
& s_{z,k+1}=s_{zk}-\sum_d\big(x_{zdk}+y_{zdk}\big)+\sum_o\big(x_{oz,k+1-\tilde t_{oz}}+y_{oz,k+1-\tilde t_{oz}}\big)+s^{\text{arr}}_{z,k+1},
\end{aligned}
\label{eq:A}
\end{equation}
where $x$ are rebalancing moves, $y$ are served passenger trips and $s$ are available drivers; $s_{z0}=s_z$ is the drivers now available in zone $z$, as under R; $s^{\text{arr}}_{zk}$ are the drivers committed to arrive in zone $z$ in interval $k$; $\tilde t_{od}=\max\{1,\operatorname{round}(t_{od}/\Delta T_p)\}$ are the travel time $t_{od}$ counted in price intervals; $\hat q_{odk}$ are the forecast requests from $o$ to $d$ in interval $k$, set for each experiment in Section~\ref{sec:setup}; $\overline{\text{fare}}_{od}$ is the fare of an average trip from $o$ to $d$ at zone $o$'s current rate; and $\bar b_{od}$ is the compensation of Eq.~\eqref{eq:compgen} for a move between the centres of $o$ and $d$, at $d$'s current shortage. 

With the definitions of two rebalancing policies and the shared ride-hailing system, the next section describes how drivers make decisions to interact with the platform, as well as how the LLM agents are set up to make those decisions.

\subsection{Driver decisions by large language models}
\label{sec:llm}

Other than the compliant driver baseline (Section~\ref{sec:results}), all experiments in this paper have drivers who make their own decisions, which are modeled by large language models (LLMs). The system status and rebalancing policies are prompted as plain language and serve as the sole input to the LLMs, which return a decision. The LLMs are not trained or rewarded to imitate human drivers, and they have no memory beyond the prompt. They are therefore a stand-in for the assistant a driver could hand the app and the disclosed rules to, and their decisions are self-interested: they maximize their own net income. The drivers of the baseline, by contrast, comply with the platform: they accept every passenger request and rebalancing offer, and never move on their own. The ideal benchmark would be a sophisticated model of how human drivers behave, but such a model, validated for drivers who act on disclosed rules, is hard to obtain. We are therefore transparent about this choice and use compliant drivers as the benchmark, which is also the behaviour that rebalancing policies are usually designed and evaluated for (Section~\ref{sec:rw-rebalancing}).

Each driver is represented by an individual LLM, or AI agent, as follows. Whenever that driver has a decision to make, the agent receives one prompt containing the currently available options together with the latest information about the system, and returns one selected option as the chosen action of the driver. The four subsections below explain the process of making decisions by AI agent: how the system is represented as a prompt (Section~\ref{sec:inputs}), how the policies are disclosed to the agent (Section~\ref{sec:disclosure}), when and what options are available to the agent, and how the agent makes decisions (Section~\ref{sec:decisions}), and lastly how different LLM models make decisions (Section~\ref{sec:models}). 

\subsubsection{System status inputs}
\label{sec:inputs}

Table~\ref{tab:prompt} summarizes the sections of the prompt that an agent receives when making a decision. \texttt{[ROLE]} gives the agent's role and objective, and \texttt{[PLATFORM RULES]} the disclosed rules. \texttt{[TIME]} and \texttt{[YOU]} describe the time and the driver's own state. \texttt{[YOUR OFFER]} and \texttt{[YOUR REQUEST]} describe the offer or request the agent is answering, and appear only then. \texttt{[MARKET]} describes the market in the driver's zone and the whole city, and \texttt{[DEMAND]} the demand in every zone. The questions follow the prompt; the options for driving to another zone give, for each zone, its fare, the compensation of an offer to it, its idle drivers, recent requests and match rate, and the time, distance and cost of getting there. Appendix~\ref{app:prompts} gives the prompts in full. The prompt is updated at every decision step, and the agent receives the latest information about the system when making a decision.

\begin{table}[htbp]
\centering
\small
\caption{Sections of the prompt an agent receives at each decision in the city model, in the order it reads them. Appendix~\ref{app:prompts} gives the full prompts of both models.}
\label{tab:prompt}
\begin{tabular}{L{3.5cm} L{10.6cm}}
\toprule
Section & Content \\
\midrule
\texttt{[ROLE]} & Role and goal \\
\texttt{[PLATFORM RULES]} & The disclosed rules (Section~\ref{sec:disclosure}); empty under N \\
\texttt{[TIME]} & Clock time and the end of the shift \\
\texttt{[YOU]} & Zone, state, share of each fare kept, own history over the last six price intervals \\
\texttt{[YOUR OFFER]} & Only when an offer is held: origin, destination zone and compensation \\
\texttt{[YOUR REQUEST]} & Only when a request is assigned: pickup and drop-off zones, trip distance and time, pickup time, fare and the driver's share, earlier rejections \\
\texttt{[MARKET]} & Own zone and city totals over the last six price intervals: requests, idle drivers, matched requests, fare, and the compensation of an offer to each zone \\
\texttt{[DEMAND]} & By zone: average fare per km and requests over the last hour, expected requests over the next hour, road distance from the driver \\
\bottomrule
\end{tabular}
\end{table}

\subsubsection{Policy disclosure}
\label{sec:disclosure}

The policy is written to the level the directive asks for. It follows Article 9(1)(b)(ii)--(iii): it names the main parameters of the platform's decisions, their relative importance and how the driver's behaviour affects the decisions, with no formulas or numbers. It is represented as a section of the prompt with six plain-language paragraphs: how the platform works over time; how the platform matches requests; rebalancing offers; how the platform decides the fares and compensation; how the drivers' actions affect the platform's decisions; and how the rebalancing works. The last paragraph describes the active rebalancing policy, which in this study is either reactive or anticipatory; all other paragraphs are always identical.

\begin{figure}[H]
\begin{rulebox}[Rebalancing paragraph under R (reactive)]
Relocation. At regular intervals, the platform counts the requests lost in each zone since its last plan. Starting from the drivers now available in each zone, it sends drivers to cover these lost requests with as little travel time as possible: drivers already in that zone cover it first, without an offer, and the rest come from other zones through relocation offers, one per lost request, while drivers are available. Lost requests include those lost because no driver was close enough and those lost after a fixed number of rejections. A request that is lost after you reject it counts among your zone's lost requests. The lost requests decide where offers go and how many; travel time decides which zones the drivers come from.
\end{rulebox}
\begin{rulebox}[Rebalancing paragraph under A (anticipatory)]
Relocation. At regular intervals, the platform forecasts the requests in each zone for the near future. Starting from the drivers now available in each zone and those on their way to one, it plans driver movements over that time to maximize its expected commission from fares minus the compensation it pays. The forecasts and the drivers in each zone decide where drivers will be short; comparing the commission a moved driver is expected to bring with the compensation decides whether to move one. Only the movements planned for now are sent as relocation offers; the plan is redone at each interval.
\end{rulebox}
\caption{The rebalancing paragraph of the platform rules under each policy. The rules call rebalancing offers ``relocation offers''.}
\label{fig:relocation}
\end{figure}

The details of shared prompts are given in Appendix~\ref{app:prompts}. With all information in the prompt, the agent can read the rules and understand how the platform works, and how its own actions affect the platform's decisions. The next section describes how the agent makes decisions based on this information.

\subsubsection{Decisions}
\label{sec:decisions}

A driver has three kinds of decision points, which are summarized in Table~\ref{tab:actions}. The first is when the driver is idle, which happens once every idle-decision interval $\Delta T_i$. The driver can either stay at its current location, or move to a different zone. If the driver chooses to move, driving cost will be incurred and the driver will not be paid for the movement. During the movement, the driver will still be idle and available to serve any incoming requests. 
The second is when a passenger request is assigned to the driver. The driver can either accept or reject the request. If the driver accepts the request, it will be committed to serve the passenger until the trip is completed. If the driver rejects the request, it can then choose either to stay at its current location or move to a different zone, which is the same as the idle decision. After rejecting a request, the driver will not be assigned any new requests until the next decision point with duration $\Delta T$.
The third is when a rebalancing offer is assigned to the driver. Similar to the request decision, the driver can either accept or reject the offer. If the driver accepts the offer, it will be committed to move to the destination zone and will not be matched with any requests during the trip. If the driver rejects the offer, it can then choose either to stay at its current location or move to a different zone, which is the same as the idle decision. After rejecting an offer, the driver will not be assigned any new offers until the next decision point with duration $\Delta T_r$.

\begin{table}[htbp]
\centering
\small
\caption{Actions at each decision point.}
\label{tab:actions}
\begin{tabular}{L{3.5cm} L{10.6cm}}
\toprule
Decision point & Actions \\
\midrule
Idle & STAY/MOVE \\
On request & ACCEPT/REJECT; if REJECT: STAY, MOVE \\
On rebalancing offer & ACCEPT/REJECT; if REJECT: STAY, MOVE \\
\bottomrule
\end{tabular}
\end{table}

\subsubsection{LLM models}
\label{sec:models}

In this study, two LLMs are tested and compared. Considering the cost of running city-scale simulations, the main agent chosen is Jev, a TypeSafe System One model \citep{typesafe2026jev}, which is significantly cost-effective for large-scale simulations. The control agent is Claude Haiku 5.5 with xhigh effort, serving as a reasoning-model benchmark. 

\paragraph*{Jev.} Jev takes plain-language prompts as input, and returns a set of probabilities over the predefined options, and the option with the highest probability is selected as the decision. 

\paragraph*{Claude Haiku 5.5.} The reasoning control is Claude Haiku 5.5 at effort \emph{xhigh}. The same input prompt is sent to the model, and it returns the selected action from the predefined options. The model is designed to have strong reasoning capabilities, and serves as a benchmark for the Jev model.

The versions of the models, their cost per API call and other details are given in Appendix~\ref{app:models}. With the two LLMs set up to make decisions, the next section describes the experiments that put the method to work.

\section{Experimental setup}
\label{sec:setup}

We evaluate the effect of disclosure on driver behaviour and market outcomes in two experiments. The first is a small three-zone model with ten drivers, which serves as a controlled environment. The second is a city-scale experiment in Manhattan, which shows how the results scale up to a realistic market.

Both experiments cross the two rebalancing policies of Section~\ref{sec:policies}, reactive (R) and anticipatory (A), with two disclosure conditions, no disclosure (N) and law-level disclosure (L), giving the four test cases of Table~\ref{tab:design}.

\paragraph*{Disclosure conditions.} The condition decides only what the \texttt{[PLATFORM RULES]} section of the prompt (Table~\ref{tab:prompt}) contains.
\begin{itemize}
\item \textbf{No disclosure (N).} The \texttt{[PLATFORM RULES]} section is empty, and every other section is identical to L. The agent knows that the platform sets fares, assigns requests and sends rebalancing offers, but not how it makes these decisions, so it can only infer them from the market state and its own history. This is the situation of today's drivers, who work out from their consequences the rules they cannot see \citep{mohlmann2021algorithmic}.
\item \textbf{Law-level disclosure (L).} The \texttt{[PLATFORM RULES]} section holds the six plain-language paragraphs of Section~\ref{sec:disclosure}. They are written to the level of Article 9(1)(b)(ii)--(iii) of the directive: the main parameters of the platform's decisions, their relative importance and how the driver's behaviour affects them, with no formulas or numbers. The first five paragraphs are the same under R and A; only the rebalancing paragraph (Figure~\ref{fig:relocation}) names the active policy.
\end{itemize}
Under both conditions the agent sees the same past demand, historical demand pattern, market state and own history, as a real driver knows them from experience. Disclosure therefore only provides extra information about the platform's decision-making process.

\begin{table}[htbp]
\centering
\small
\caption{The $2\times2$ design: each test case pairs a rebalancing policy with a disclosure condition.}
\label{tab:design}
\begin{tabular}{L{3.0cm} L{5.4cm} L{5.4cm}}
\toprule
 & N: no disclosure & L: law-level disclosure \\
\midrule
R: reactive & R$\times$N: offers cover lost requests; rules not shown & R$\times$L: offers cover lost requests; rules shown, with the R rebalancing paragraph \\
A: anticipatory & A$\times$N: offers follow the forecast plan; rules not shown & A$\times$L: offers follow the forecast plan; rules shown, with the A rebalancing paragraph \\
\bottomrule
\end{tabular}
\end{table}

\paragraph*{Test cases.} R$\times$N and A$\times$N are the platform as drivers know it today, running reactive or anticipatory rebalancing behind rules they cannot see; R$\times$L and A$\times$L are the same platforms once the directive applies. Comparing N with L under one policy measures the effect of disclosure, comparing R with A under one condition measures the effect of the policy, and the difference between the two disclosure effects, the policy $\times$ disclosure interaction (Section~\ref{sec:metrics}), shows whether the effect of disclosure depends on the policy. Within an experiment the four cells share everything else, the requests, the start positions and the random seeds, so each comparison changes one factor at a time. The two experiments, a three-zone small model (Section~\ref{sec:setup-small}) and Manhattan (Section~\ref{sec:setup-city}), are described next; their parameter values are listed in Table~\ref{tab:params} (Appendix~\ref{app:params}).

\subsection{Small model}
\label{sec:setup-small}

To understand the effect of disclosure on driver behaviour and market outcomes, we first test the four cases in a small model with three zones and ten drivers. The small model is a controlled environment to understand how disclosure affects driver behaviour and market outcomes, before the city-scale experiment.

\begin{figure}[htbp]
\centering
\includegraphics[width=\linewidth]{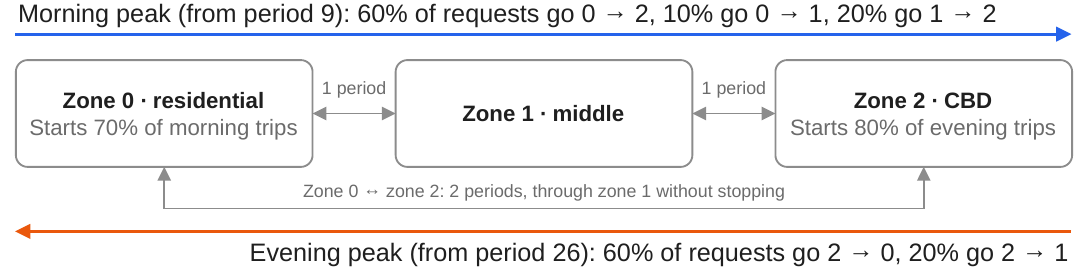}
\caption{The network of the small model, with the travel times between zones and the main flows of the two demand peaks.}
\label{fig:smallmodel}
\end{figure}

\paragraph*{Network and demand.} 
There are three zones, connected in a line: residential (zone 0), middle (zone 1) and the central business district (zone 2). Travelling between neighbouring zones takes one period, and between the two end zones two periods, passing through the middle zone without stopping. Demand is tidal, towards zone 2 in the morning and the reverse in the evening, with mean requests per period
\begin{equation}
\lambda(t)=\lambda_0+\lambda_1\Big[\exp\!\Big(-\big(\tfrac{t-t^{\text{am}}}{\sigma}\big)^{2}\Big)+\exp\!\Big(-\big(\tfrac{t-t^{\text{pm}}}{\sigma}\big)^{2}\Big)\Big],\qquad \lambda_{od}(t)=\lambda(t)\,w_{od}(t),
\label{eq:demand}
\end{equation}
where $t$ is the period of the day, with a morning peak at period $t^{\text{am}}$ and an evening peak at $t^{\text{pm}}$. Every request leaves its zone. The origin--destination shares $w_{od}(t)$ send most morning requests towards the business district, and reverse in the evening, with a linear shift in between. The demand is pre-defined and frozen for all runs through a Poisson draw. Figure~\ref{fig:smallmodel} represents the network visually.

\paragraph*{Platform.} The three decision intervals ($\Delta T$, $\Delta T_p$ and $\Delta T_r$) described in Section~\ref{sec:system} are all one period long in the small model, so the platform decides matches, fares, rebalancing offers and its compensation at the same time. A driver has no position within its zone, so in Eq.~\eqref{eq:compgen} $\ell_{vd}=\ell_{od}$ for a driver in zone $o$, and the surge rate $\kappa_{vd}$ is the same for every move: an offer pays a fixed amount per unit of distance plus a flat amount that rises with the destination's shortage. 
Drivers waiting in a zone form a queue in descending order of time waited in this zone, where requests and rebalancing offers are assigned from the front of the queue. A driver who rejects a request will be placed at the back of the queue, and gets no further request until the next period. 
The idle drivers $\bar s_z$ of Eq.~\eqref{eq:shortage} are those still idle after requests are assigned, without considering the drivers who rejected a request and drivers arriving on their own. 
Both rebalancing policies plan at every period, and anticipatory policy looks $H$ periods ahead, with its forecast $\hat q_{odk}$ set to the mean demand $\lambda_{od}(t)$ of Eq.~\eqref{eq:demand} in each of those periods.

\paragraph*{Drivers.} The system operates with ten drivers, which are all AI agents. There are three different initializations of the drivers' start positions, with each zone to start with 4 drivers once, and 3 drivers in the other two zones. Each initialization is repeated five times with different random seeds. The AI agent sees past demand and the historical pattern, and makes decisions at every period. The drivers' actions are self-interested, and they aim to maximize their own net income.

\subsection{City model}
\label{sec:setup-city}

A more realistic experiment is run on the Manhattan road network, with 500 drivers. It shows how the results scale up to a realistic market.

\begin{figure}[htbp]
\centering
\includegraphics[width=\linewidth]{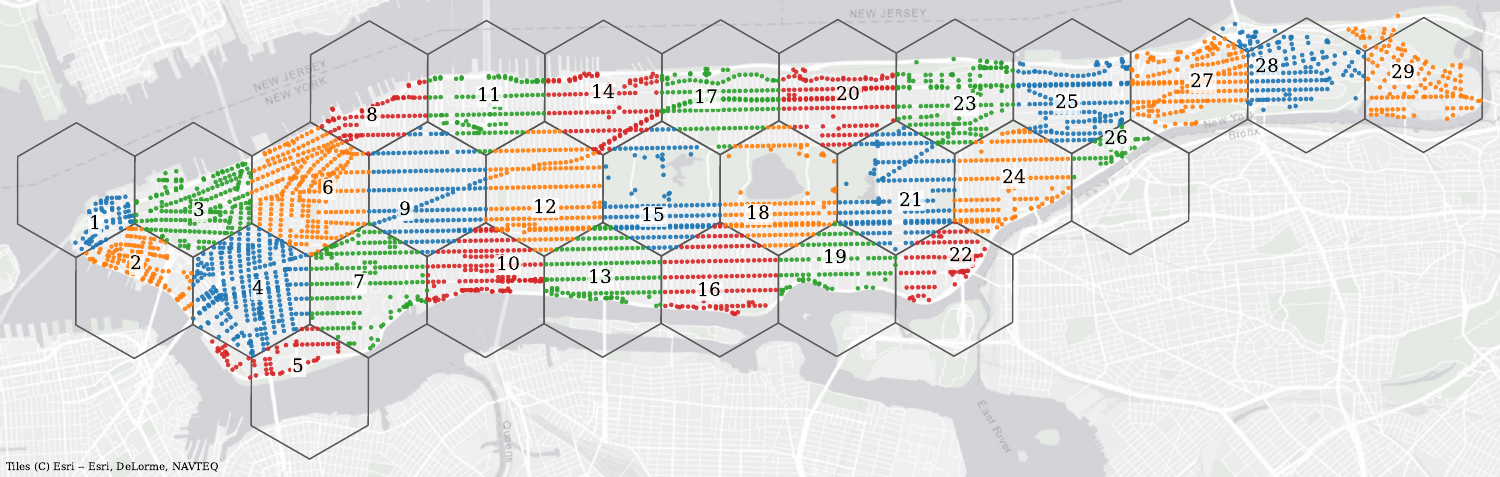}
\caption{The Manhattan network and its 29 hexagonal zones. The map is rotated 60 degrees clockwise.}
\label{fig:manhattan}
\end{figure}

\paragraph*{Network and demand.} The Manhattan network is divided into 29 hexagonal zones of 1 km radius (Figure~\ref{fig:manhattan}); travel times and distances follow the roads, and distances $\ell_{od}$ are road kilometres. Passenger requests are sampled with a share $\rho$ from the yellow-taxi records of March 2016, and the test day is Wednesday 23 March 2016. This day is chosen as it is a workday without large events or major road closures, and its demand is within 3\% of the March Monday-to-Thursday median for the whole day and both peaks. The training days are 1--3 and 7--10 March; their mean demand at the same time of day, scaled by the same share $\rho$, is A's forecast $\hat q_{odk}$.

\paragraph*{Platform.} At every decision step, each waiting request is matched to an idle driver who can reach it within $t_w^{\max}$ of the request, preferring the shortest pickup time; a rejected request goes to another driver who can still reach it in time, and it is lost after the maximum number of rejections or when no such driver is left. Fares are updated every price interval at a base rate $p$ per km; the idle drivers $\bar s_z$ of Eq.~\eqref{eq:shortage} are 
average number of idle drivers in the zone over the interval. The driving cost of Eq.~\eqref{eq:cost} is $\beta_\ell$ per km plus $\beta_T$ per hour online. In the compensation of Eq.~\eqref{eq:compgen}, both payments are a share $\alpha$ of the destination's fare rate, $b_0=\alpha\,p$ and $\kappa_{vd}=\alpha\,p\,\ell_{vd}$, so an offer to driver $v$ for zone $d$ pays
\begin{equation}
b_{vd}=\alpha\,p\,(1+g_d)\,\ell_{vd},
\label{eq:comp}
\end{equation}
where $\ell_{vd}$ is the road distance from the driver's position to the zone's centre; A's plan uses $\bar b_{od}$, with the distance $\ell_{od}$ between the zone centres in place of $\ell_{vd}$. Rebalancing is planned every planning interval $\Delta T_r$, and A looks $H$ price intervals $\Delta T_p$ ahead. A driver moving on their own stays available along the way and can be matched or receive an offer in any zone it passes through.

\paragraph*{Drivers and design.} The fleet contains 500 drivers, which are all represented by AI agents, except in the mixed fleets of Section~\ref{sec:res-mixed}, where 100 or 250 of them are AI agents and the others comply. The drivers are initialized at random positions, remain online during the entire simulation, and make own decisions at every decision step. A warm-up period of 1 hour is run before the actual simulation, where every driver complies to the system during the warm-up. A compliant baseline where all drivers comply is also run with same random seeds in pair to serve as a benchmark. Table~\ref{tab:city_exp} summarizes the experiment setup.

\begin{table}[htbp]
\centering
\small
\caption{Set-up of the city experiment. Parameter values are in Table~\ref{tab:params} (Appendix~\ref{app:params}).}
\label{tab:city_exp}
\begin{tabular}{L{4.2cm} L{10.0cm}}
\toprule
Item & Setting \\
\midrule
Network & Manhattan road network with 4091 nodes, divided into 29 hexagonal zones \\
Test day & Wednesday 23 March 2016, 4:00--24:00 \\ 
Warm-up & 3:00--4:00 (all drivers are compliant) \\
Total Request & \num{52461}\\
Fleets & 500 AI Drivers; In mixed fleets: 100 or 250 AI drivers and the others complying \\  
\bottomrule
\end{tabular}
\end{table}

\subsection{Comprehension test}
\label{sec:comprehension}

Before conducting the experiments, a comprehension test is run to check whether the AI agent understands the platform and rules. The test is conducted under the L condition with all relevant information provided as prompt input, without actual system status. A pre-defined set of multiple-choice questions is asked to the AI agent, and its answers are compared to the correct answers. The test is designed to cover both how the system works and what the disclosed policy says, and it is run for both policies, R and A, and for both models, Jev and Claude Haiku 5.5.

The questions are divided into two groups. Group (a) covers how the system works: the direction of the fare response to idle drivers, rejections, committed arrivals and demand; which demand the fare uses; when compensation is locked and paid; what happens to requests after a rejection; and when fares are settled. Group (b) covers the disclosed policy: what decides where offers go, lost requests under R and the forecast under A. Extra questions are asked for the city-scale model, which ask how compensation is set and how it moves with the destination's shortage, declined offers, rejecting and staying, moving on one's own, the regular idle choice and when fares are paid. The full questions are listed in Appendix~\ref{app:questions}.

Section~\ref{sec:res-agents} reports the results of the test, and the next subsection describes the benchmarks and metrics used to evaluate the effect of disclosure on driver behavior and market outcomes.

\subsection{Metrics of evaluation}
\label{sec:metrics}

To evaluate the effect of disclosure on driver behaviour and market outcomes, several benchmarks and metrics are defined and used in the experiments. There are three groups of benchmarks, which correspond to the three parties in the market: the driver, the platform, and the users. The metrics are defined as follows.

\paragraph*{Driver.} As the main focus of this study, the AI driver is the only party whose behaviour is affected by disclosure. The driver's actions are self-interested, and they aim to maximize their own net income. The driver's income is its net income over the simulated day: its share $1-\theta$ of the fares of the requests it serves, plus the compensation of the rebalancing offers it completes, minus its driving and time cost $C_v$ of Eq.~\eqref{eq:cost},
\begin{equation}
I_v=\sum_{r\in\mathcal{R}_v}(1-\theta)\,\text{fare}_{o_r d_r}+\sum_{j\in\mathcal{O}_v} b_{v d_j}-C_v,
\label{eq:income}
\end{equation}
where $\mathcal{R}_v$ is the set of requests driver $v$ serves with fares from Eq.~\eqref{eq:fare}, and $\mathcal{O}_v$ the set of rebalancing offers it completes, each paying the compensation $b_{vd_j}$ for its destination $d_j$ (Eq.~\eqref{eq:compgen}).

\paragraph*{Platform.} The platform's profit is its commission $\theta$ of the fares of the requests served, minus the compensation it pays to drivers for rebalancing offers,
\begin{equation}
\Pi=\sum_{v\in V}\Big(\sum_{r\in\mathcal{R}_v}\theta\,\text{fare}_{o_r d_r}-\sum_{j\in\mathcal{O}_v} b_{v d_j}\Big),
\label{eq:profit}
\end{equation}
summed over the fleet $V$ and the simulated day, with $\mathcal{R}_v$, $\mathcal{O}_v$ and $b_{vd_j}$ as in Eq.~\eqref{eq:income}. 

\paragraph*{User.} The users' benefits are represented in two parts. The first is the service rate, which quantifies the service quality. The second is the served user cost, which combines the waiting time and the fare of each served request $r$ into one monetary cost,
\begin{equation}
C_{u,r}=\beta_w\,t_{w,r}+\text{fare}_{o_r d_r},
\label{eq:usercost}
\end{equation}
where $t_{w,r}$ is the time from the request to the pickup, $\beta_w$ the value of waiting time, which converts it into money, and the fare is that of Eq.~\eqref{eq:fare}, fixed when the request is made. 

\section{Results}
\label{sec:results}

In both the small model and the city, the results of the experiments where all drivers are AI agents are compared against the benchmark in which every driver complies with the platform, to evaluate the effect of disclosure on driver behaviour and market outcomes under two different policies. Section~\ref{sec:res-agents} first checks that the AI agents understand the platform and rules, and compares the two LLMs in the small model (Table~\ref{tab:small}). Section~\ref{sec:res-city} then compares the two disclosure conditions, N and L, in the city (Table~\ref{tab:cityresults}), first under the reactive policy and then under the anticipatory policy, and Section~\ref{sec:res-RA} compares how the two policies respond to disclosure. Section~\ref{sec:res-mixed} then lets only some of the city's drivers use an AI agent.

\subsection{The AI drivers: comprehension and behaviour in different LLMs}
\label{sec:res-agents}

\paragraph*{Comprehension test.} Before presenting the results of the experiments, we first check if the AI agents understand the platform and rules. The comprehension test is run for both models, Jev and Claude Haiku 5.5, under both policies, R and A, with the same prompt input as in the experiments. The results are summarized in Table~\ref{tab:comprehension}. Both models answer all questions correctly, with confidence level at least 0.92 for Jev. This includes the question on what decides where rebalancing offers send drivers, whose correct answer differs between R and A, so the agents can tell which policy is active. With the comprehension level ensured, we then explain the experiment results in the small model.

\begin{table}[htbp]
\centering
\small
\caption{Comprehension test under disclosure L: rule text only, no market state, multiple choice.}
\label{tab:comprehension}
\begin{tabular}{L{2.2cm} L{3.6cm} c c c}
\toprule
Scale & Model & R & A & Minimum confidence \\
\midrule
Small model & Jev & 12/12 & 12/12 & 0.92 \\
 & Claude Haiku 5.5 (xhigh) & 12/12 & 12/12 & -- \\
City & Jev 1.13.0 & 17/17 & 17/17 & 0.94 \\
\bottomrule
\end{tabular}
\end{table}

\paragraph*{Small model.} We first evaluate the two policies under the disclosure conditions N and L. Results are shown in Table~\ref{tab:change}. In both policies R and A, AI agents are able to utilize the disclosed information to make better decisions, which in turn increases their own income, as prescribed in their objectives. Under both LLMs, the drivers' own moves, which are the unpaid movements other than serving the rebalancing offers from the system, are reduced under L. This reduction reflects the drivers' better understanding of the platform's rebalancing policies, which allows them to make more informed decisions and avoid unnecessary movements. With the fleet understanding the platform better, the service rate is improved in all test cases. However, the platform's profit does not follow the same universal trend across the four test cases. While it increases in the other three test cases, including R with Haiku, the platform's profit under Jev and R decreases (the bold number in Table~\ref{tab:change}). To better understand why such an outlier occurs, we further investigate the relationship between the drivers' own moves and the platform's profit.

\begin{table}[htbp]
\centering
\caption{Small model results with all ten drivers as AI agents: means over all drivers and runs. Brackets give the change against compliant benchmark; acc.\ and decl.\ mean accepted and declined.}
\label{tab:small}
\begin{subtable}{\linewidth}
\centering
\scriptsize
\setlength{\tabcolsep}{2.5pt}
\caption{Results per cell.}
\label{tab:allai}
\begin{tabular}{l l c c c c c c c c c}
\toprule
 & & & & & \multicolumn{2}{c}{Offers} & & & \multicolumn{2}{c}{Requests} \\
\cmidrule(lr){6-7}\cmidrule(lr){10-11}
Cell & Model & \makecell{Drivers'\\Income ($I_v$) } & \makecell{Platform\\Profit ($\Pi$)} & \makecell{Service\\rate (\%)} & acc. & decl. & \makecell{Rebalancing\\Compensation} & \makecell{Own\\moves} & acc. & decl. \\
\midrule
R $\times$ N & Jev & 4.90 ($-$0.14) & 17.80 ($+$10.00) & 41.5 ($-$9.4) & 2.7 & 1.1 & 0.69 & 9.4 & 7.4 & 0.0 \\
 & Haiku (xhigh) & 4.08 ($-$0.96) & 13.96 ($+$6.17) & 32.3 ($-$18.5) & 2.9 & 1.1 & 0.76 & 15.0 & 5.8 & 0.0 \\
R $\times$ L & Jev & 5.02 ($-$0.01) & 15.08 ($+$7.28) & 51.8 ($+$1.0) & 3.2 & 0.4 & 0.86 & 4.9 & 9.3 & 0.0 \\
 & Haiku (xhigh) & 5.41 ($+$0.37) & 20.67 ($+$12.87) & 48.2 ($-$2.6) & 2.5 & 0.7 & 0.66 & 10.9 & 8.6 & 0.0 \\
\midrule
A $\times$ N & Jev & 4.77 ($-$0.94) & 9.86 ($-$0.80) & 43.7 ($-$15.5) & 5.4 & 1.3 & 1.21 & 6.3 & 7.8 & 0.0 \\
 & Haiku (xhigh) & 3.34 ($-$2.37) & 0.93 ($-$9.73) & 29.7 ($-$29.5) & 6.4 & 1.3 & 1.45 & 12.9 & 5.3 & 0.0 \\
A $\times$ L & Jev & 5.53 ($-$0.18) & 11.69 ($+$1.03) & 57.8 ($-$1.4) & 6.7 & 0.9 & 1.28 & 1.2 & 10.4 & 0.0 \\
 & Haiku (xhigh) & 5.02 ($-$0.69) & 9.10 ($-$1.56) & 48.0 ($-$11.2) & 6.6 & 1.6 & 1.35 & 6.4 & 8.6 & 0.1 \\
\bottomrule
\end{tabular}
\end{subtable}

\vspace{8pt}
\begin{subtable}{\linewidth}
\centering
\footnotesize
\caption{Change from N to L (L minus N, from panel~a).}
\label{tab:change}
\begin{tabular}{l cccc}
\toprule
Change from N to L & Jev, R & Haiku, R & Jev, A & Haiku, A \\
\midrule
Driver income ($I_v$) & $+$0.13 & $+$1.34 & $+$0.76 & $+$1.68 \\
Driver own moves & $-$4.5 & $-$4.1 & $-$5.1 & $-$6.5 \\
Service rate (\%) & $+$10.4 & $+$15.9 & $+$14.1 & $+$18.3 \\
Platform profit ($\Pi$) & \textbf{\boldmath$-$2.7} & $+$6.7 & $+$1.8 & $+$8.2 \\
\bottomrule
\end{tabular}
\end{subtable}
\end{table}

\begin{figure}[htbp]
\centering
\begin{subfigure}[t]{0.45\linewidth}
\includegraphics[width=\linewidth]{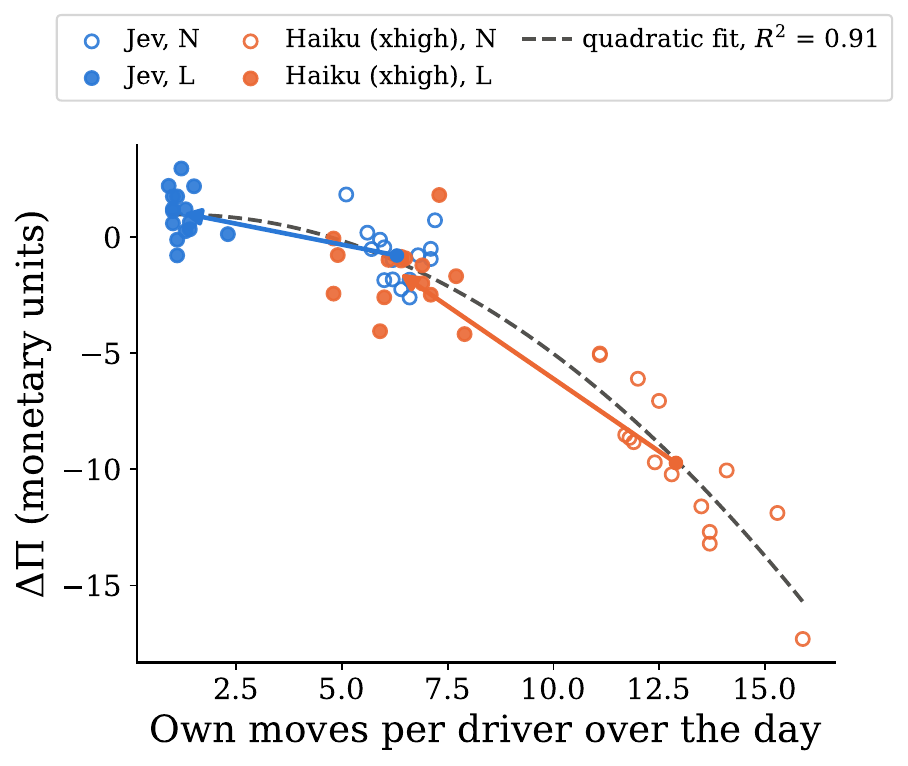}
\caption{Policy A}
\end{subfigure}
\begin{subfigure}[t]{0.45\linewidth}
\includegraphics[width=\linewidth]{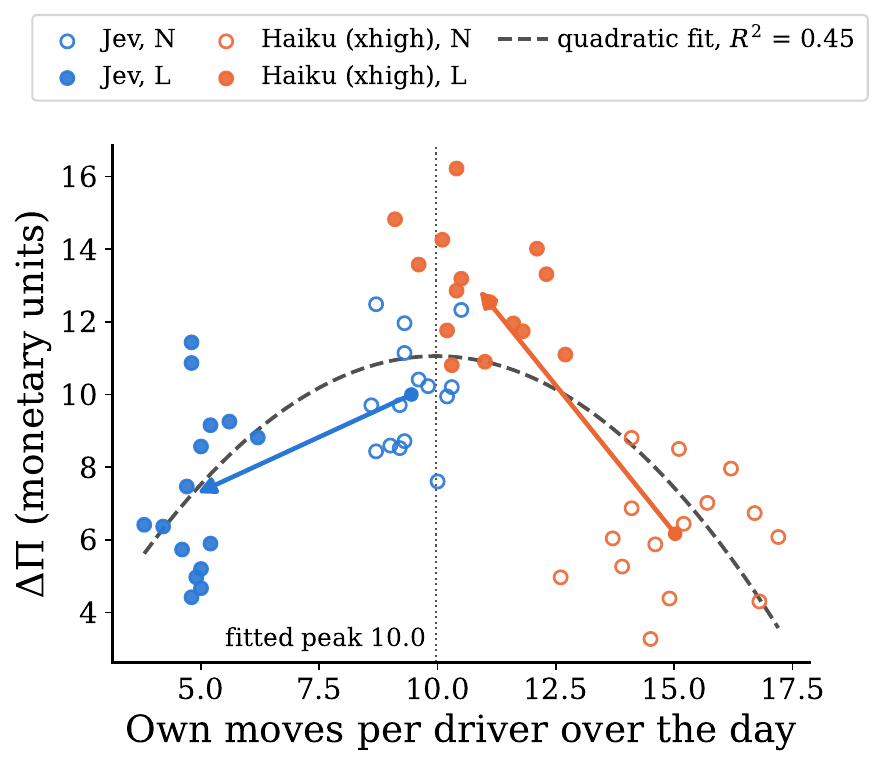}
\caption{Policy R}
\end{subfigure}
\caption{Own moves per driver against the change in platform profit relative to everyone complying ($\Delta\Pi$) in the small model with all ten drivers AI, one dot per run. The fit is over both models and both conditions; arrows join each model's N and L cell means.}
\label{fig:movescurve}
\end{figure}

In Figure~\ref{fig:movescurve}, we plot the average number of drivers' own moves against the change in platform profit relative to everyone complying. Firstly, we observe that in all test cases, Jev tends to move less than Haiku. This should be attributed to the nature of the LLM, since both of the models show strong comprehension of the system. Additionally, we observe different patterns of the relationship between the drivers' own moves and the platform's profit under different policies. Under policy A, $\Pi$ decreases as the drivers move on their own more (Figure~\ref{fig:movescurve}a). This is because A is a reasonably good rebalancing policy in the small model: its plan already sends drivers to where demand is forecast, and with few own moves the platform's profit is close to its level with every driver complying. An own move therefore tends to take a driver away from where A's plan needs it, and the more the drivers move on their own, the lower the profit. This results in the consistent increase in the platform's profit under both LLMs when information is disclosed (from N to L) in Table~\ref{tab:change}.

However, under policy R, there is a peak in the platform's profit when the drivers move around 10 times on their own (Figure~\ref{fig:movescurve}b). The peak arises because R is not a strong rebalancing policy in the small model. It reacts only to requests that have already been lost, and with AI drivers the platform's profit is above its level with every driver complying in every run. Up to about 10 moves per driver, the drivers' own moves therefore make up for what R's offers miss; beyond that, they take drivers away from requests they could serve, and the profit falls. The peak explains the difference in the direction of the platform's profit change between the two LLMs. Since the Jev model naturally moves less than Haiku, the number of own moves of Jev under N lies around the peak of the curve, and when information is disclosed, the drivers move less, which results in a decrease in the platform's profit. On the other hand, Haiku moves more than Jev under N, and the number of own moves of Haiku under N lies on the right side of the peak, and when information is disclosed, the drivers move less, which results in an increase in the platform's profit.

Thus, the two LLMs behave in a similar way when information is disclosed under both policies. Both of them show strong comprehension of the system, and consistently make similar decisions in operations. Since Jev responds to disclosure similarly to a larger reasoning LLM, Haiku, while with cost less than 10\% per call (Table~\ref{tab:models}), we deploy Jev as the AI driver in the following city model experiments. To better understand the effect of disclosure on driver behaviour and market outcomes, we next present the results in the city-scale experiment, which is more realistic and complex.

\FloatBarrier
\subsection{The city experiment: effect of disclosure under each policy}
\label{sec:res-city}

In the city, under the compliant benchmark (Table~\ref{tab:citybaseline}), R and A perform similarly from the platform's perspective, with the same platform profit (136.5k against 136.6k USD a day). R serves more passengers (57.8\% against 51.2\%) and sends more rebalancing offers, each with a higher compensation (2508 against 1016 a day, at 6.2 against 3.2 USD per offer). From the driver's perspective, complying with the platform results in a lower income under A (520 against 592 USD a day) and a lower utilization rate (54.5\% against 61.3\%). From the user's perspective, R is better, as it serves more passengers, while those who are served receive a similar service under the two policies, with the same mean wait and similar fares. With the benchmark established, we then evaluate the effect of disclosure on driver behaviour and market outcomes under each policy in turn, starting with R.

\begin{table}[htbp]
\centering
\small
\caption{City experiment benchmark: all drivers are compliant.}
\label{tab:citybaseline}
\begin{tabular}{L{6.4cm} cc}
\toprule
 & R & A \\
\midrule
Service rate (\%) & $57.8$ & $51.2$ \\
Utilization (\%) & $61.3$ & $54.5$ \\
Platform profit (USD/day) & $\num{136500}$ & $\num{136600}$ \\
Commission (USD/day) & \num{152100} & \num{139900} \\
Compensation paid (USD/day) & \num{15600} & \num{3270} \\
Rebalancing offers accepted & \num{2508} & \num{1016} \\
Driver net income, mean (10th--90th percentile), USD & 592 (513--672) & 520 (284--685) \\
Mean passenger wait (min) & 2.9 & 2.9 \\
Mean fare (USD) & 16.7 & 17.3 \\
\bottomrule
\end{tabular}
\end{table}

\subsubsection{Reactive rebalancing policy (R)}
\label{sec:res-R}

The city results are given in Table~\ref{tab:cityresults}. As in the small model, the drivers respond to disclosure by taking the platform's rebalancing offers: the share they accept rises from 22\% to 75\% (Table~\ref{tab:citychange}). Passenger requests are rarely rejected even under N, about 50 a day across the whole fleet, and almost never under L. On the other hand, the main difference compared to the small model is the drivers' own moves. Unlike in the small model, where the agents move less on their own once they know the rules, in the city these movements hardly change ($-1.0$ per driver). Note that these movements are infrequent, at about eight per driver over the day against some 60 passenger trips (Table~\ref{tab:cityallai}). Disclosure therefore acts through the rebalancing offers. Under R, driver income falls by 2.8 USD per driver from N to L, the service rate by 0.2 points and platform profit by 5.4k USD a day, or 3.7\% (Table~\ref{tab:citychange}).

\begin{table}[htbp]
\centering
\caption{City results with all 500 drivers as Jev agents.}
\label{tab:cityresults}
\begin{subtable}{\linewidth}
\centering
\scriptsize
\setlength{\tabcolsep}{2.5pt}
\caption{Results per cell. Income, offers, compensation (USD), own moves and requests are means per driver over the day; $\Pi$ is in thousand USD a day. Brackets give the change against compliant benchmark.}
\label{tab:cityallai}
\begin{tabular}{l c c c c c c c c c}
\toprule
 & & & & \multicolumn{2}{c}{Offers} & & & \multicolumn{2}{c}{Requests} \\
\cmidrule(lr){5-6}\cmidrule(lr){9-10}
Cell & \makecell{Drivers'\\Income ($I_v$) } & \makecell{Platform\\Profit ($\Pi$)} & \makecell{Service\\rate (\%)} & acc. & decl. & \makecell{Rebalancing\\Compensation} & \makecell{Own\\moves} & acc. & decl. \\
\midrule
R $\times$ N & 593.5 ($+$2.6) & 148.04 ($+$11.54) & 59.1 ($+$1.1) & 2.0 & 7.1 & 16.01 & 8.9 & 62.0 & 0.1 \\
R $\times$ L & 590.7 ($-$0.2) & 142.62 ($+$6.12) & 59.0 ($+$1.0) & 3.9 & 1.3 & 22.88 & 7.9 & 61.9 & 0.0 \\
\midrule
A $\times$ N & 570.6 ($+$50.9) & 150.96 ($+$14.09) & 57.7 ($+$6.1) & 1.2 & 5.3 & 4.15 & 7.8 & 60.5 & 0.1 \\
A $\times$ L & 575.3 ($+$55.5) & 151.29 ($+$14.43) & 58.0 ($+$6.4) & 1.8 & 0.7 & 5.39 & 8.0 & 60.9 & 0.0 \\
\bottomrule
\end{tabular}
\end{subtable}

\vspace{8pt}
\begin{subtable}{\linewidth}
\centering
\footnotesize
\caption{Change from N to L (L minus N), with its sources below the line. VKT is the whole fleet's; the synchrony index and own moves beyond the plan are those of Figures~\ref{fig:cityherding} and~\ref{fig:cityfree}e.}
\label{tab:citychange}
\begin{tabular}{l cc}
\toprule
Change from N to L & R & A \\
\midrule
Driver income ($I_v$, USD) & $-$2.8 & $+$4.7 \\
Driver own moves & $-$1.0 & $+$0.1 \\
Service rate (\%) & $-$0.2 & $+$0.3 \\
Platform profit ($\Pi$, k USD) & \textbf{\boldmath$-$5.42} & $+$0.33 \\
\midrule
Offers accepted (\% points) & $+$52 & $+$54 \\
Compensation paid (k USD) & $+$3.43 & $+$0.62 \\
Commission (k USD) & $-$1.99 & $+$0.95 \\
Fare per trip (\%) & $-$1.0 & $+$0.1 \\
VKT (\%) & $+$0.6 & $+$1.9 \\
Synchrony index, idle choices & $+$0.026 & $+$0.026 \\
Own moves beyond the plan (share) & $+$0.14 & $+$0.14 \\
\bottomrule
\end{tabular}
\end{subtable}

\end{table}

\begin{figure}[htbp]
\centering
\includegraphics[width=\linewidth]{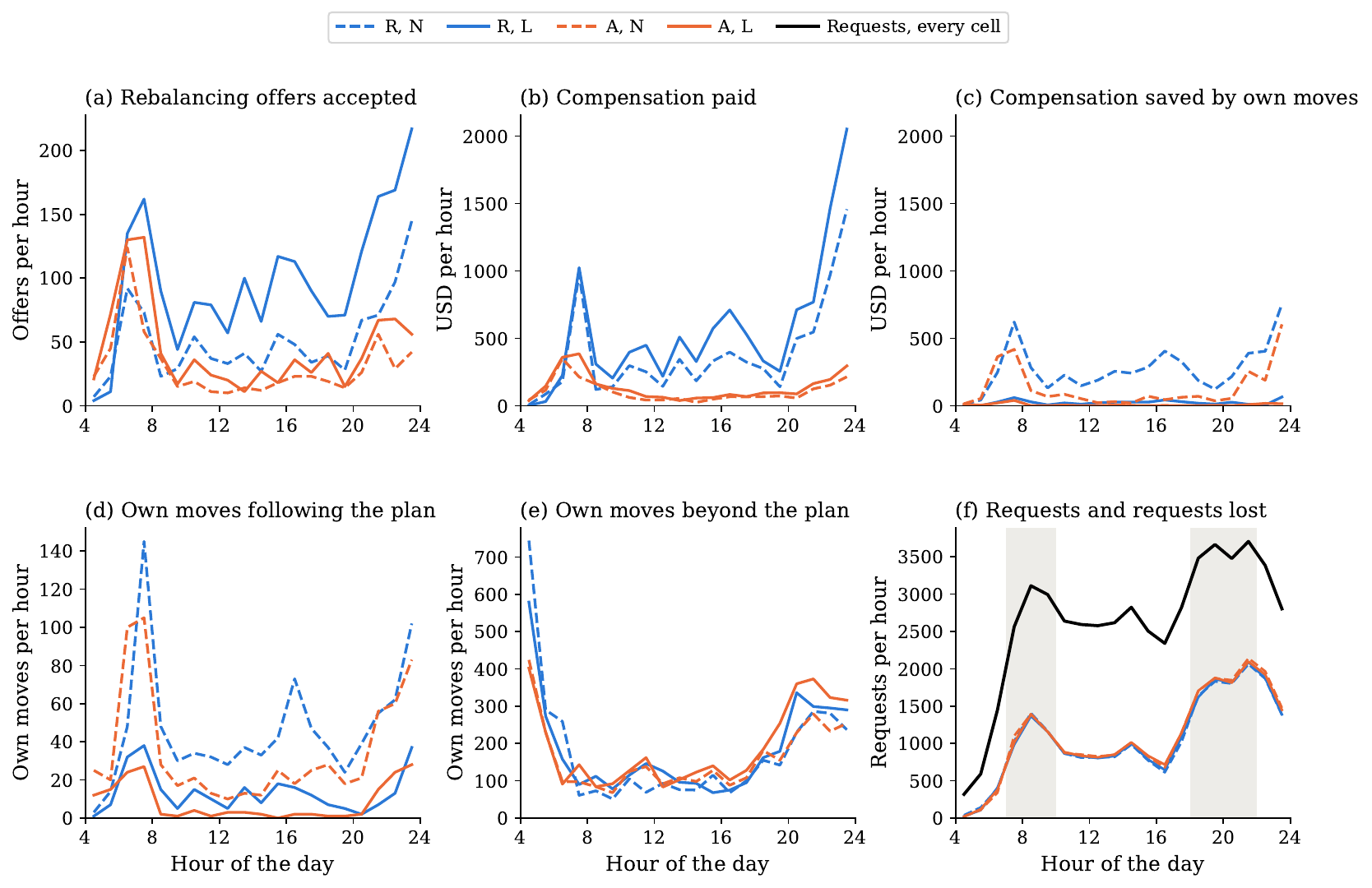}
\caption{Offers, compensation, own moves and demand by hour of the day in the city with all 500 drivers AI. In (c), an own move saves the compensation of the declined offer it stands in for; in (d) and (e), the plan is the platform's rebalancing plan. Shading marks the demand peaks.}
\label{fig:cityfree}
\end{figure}

The loss arises because R does well only while drivers decline its offers. R sends more offers than A, which are also more expensive (Table~\ref{tab:cityallai}). With every driver complying, R pays for every move it plans with 15.6k USD a day (Table~\ref{tab:citybaseline}). Under N the drivers do not know the rule behind the offers and decline about 78\% of them, but many then move on their own, often to where R meant to send a paid driver (about 930 own moves a day follow R's plan; Figure~\ref{fig:cityfree}d). R thus gets much of its planned repositioning for free: the AI drivers beat full compliance under R by 11.5k USD a day, and 66\% of that (7.6k) is compensation they declined. Once the rules are disclosed, the drivers accept 75\% of the offers, and the moves that used to be free become paid offers. With compensation rising by 3.4k USD a day, R now pays for repositioning it used to get for free. Moerover, the paid moves do not bring extra commission. Fares fall by 6.6k USD a day, and commission by 2.0k, through slightly fewer and shorter trips and a slightly lower fare rate where committed drivers lower a zone's shortage (Eq.~\eqref{eq:shortage}).

We next follow the drivers' behaviour and its cost through the day (Figure~\ref{fig:cityfree}). Demand has two peaks, in the morning (7:00--10:00) and in the evening (18:00--22:00) (Figure~\ref{fig:cityfree}f). Accepted offers (Figure~\ref{fig:cityfree}a) rise with these peaks: under both policies they peak as the morning demand builds (6:00--7:00) and again late in the evening, from 20:00 to the end of the day, and disclosure raises them at almost every hour. Compensation (Figure~\ref{fig:cityfree}b) follows the same two peaks, and R pays more than A at almost every hour: from 20:00, R $\times$ L accepts 120 to 220 offers an hour and pays up to about 2k USD an hour, while A's compensation stays below about 400 USD an hour all day under both conditions. Own moves per hour (Figure~\ref{fig:cityfree}d and e) follow the same pattern in the four test cases: high at the start of the day, low and flat through the middle of the day, and rising again in the evening. Under R, own moves fall under L only in the early morning, when the drivers take R's offers instead; in the evening they do not fall at all, while R pays for many more accepted offers. Hour by hour, then, disclosure raises what R pays, most of all at the demand peaks, without the drivers moving much less on their own. Almost half of R's extra compensation under L is paid after 20:00, late in the evening peak and after it. This follows from R's design: its plan sends paid drivers to the zones that lost requests (Section~\ref{sec:R}), the evening peak leaves many lost requests behind (Figure~\ref{fig:cityfree}f), and so R keeps paying drivers to move to those zones as the peak ends. By the time they arrive, few new requests arise there, and the compensation does not turn into commission. Under N the drivers declined most of these late offers; under L they accept them, and the waste grows. The burst of compensation at 7:00 differs: it comes while demand is still rising, and it appears under N and L alike.

\begin{figure}[htbp]
\centering
\includegraphics[width=0.8\linewidth]{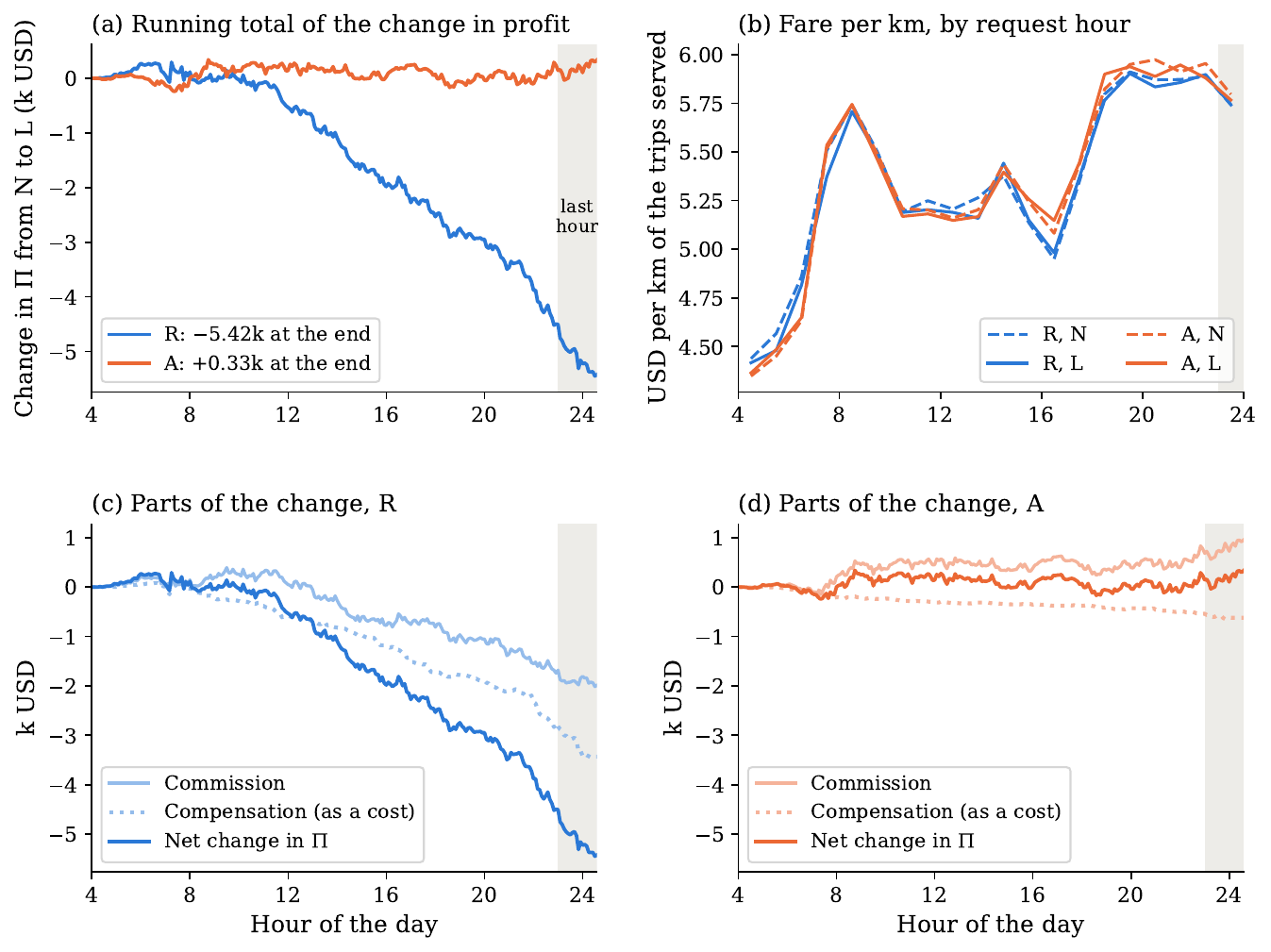}
\caption{Change in platform profit $\Pi$ from N to L through the day in the city with all 500 drivers AI, with its parts and the fare per km of the trips served. The curves run past midnight.}
\label{fig:citytiming}
\end{figure}

Figure~\ref{fig:citytiming} follows the change in platform profit from N to L through the day and shows that the late compensation brings no commission back. Under R, the running total stays close to zero until about 10:00 and then falls for the rest of the day, fastest after 21:00, to $-5.4$k USD (Figure~\ref{fig:citytiming}a). Both of its parts lower the profit (Figure~\ref{fig:citytiming}c): the extra compensation grows all day, most steeply after 21:00, while the commission, instead of rising with it, falls slowly from late morning and ends 2.0k USD lower. The fare per km of the trips served follows the two demand peaks and differs little between N and L at any hour, under either policy (Figure~\ref{fig:citytiming}b).

R's rule therefore works well only while drivers do not understand the policy. Disclosure does not break it, as R $\times$ L still beats full compliance by 4.5\%, but it removes the subsidy that the drivers' free moves gave it. We next evaluate the anticipatory policy in the same way.

\subsubsection{Anticipatory rebalancing policy (A)}
\label{sec:res-A}

Table~\ref{tab:cityresults} also gives the results under A. In the city, the drivers respond to disclosure under A as they do under R: the share of offers they accept rises from 19\% to 73\%, and they stop rejecting passenger requests (Table~\ref{tab:citychange}), while their own moves hardly change ($+0.1$ per driver). The outcomes, however, differ from R: from N to L, driver income rises by 4.7 USD per driver, the service rate by 0.3 points and platform profit by 0.3k USD a day, or 0.2\%. A is spared R's loss because its offers are few and cheap, and sent only where a moved driver is forecast to bring more commission than it costs: under L its compensation rises by 0.6k and its commission by 0.9k. Hour by hour, disclosure raises A's accepted offers at most hours, as it does R's, and the drivers' own moves shift from following A's plan to going beyond it (Figure~\ref{fig:cityfree}a, d and e), yet A's compensation stays below about 400 USD an hour all day under both conditions (Figure~\ref{fig:cityfree}b). Through the day, the extra compensation grows slowly to 0.6k USD and, from about 8:00, the commission stays above its level under N, ending 0.9k USD higher, so the running total stays close to zero all day (Figure~\ref{fig:citytiming}d). Disclosure thus changes the drivers' behaviour under A as it does under R, but A loses nothing from it.

With the effects of disclosure understood, we return to the compliant benchmark. Against full compliance, the AI drivers raise the platform's profit in every cell (Table~\ref{tab:cityallai}). Under A the gain reaches every party: drivers earn more, the platform earns more and more passengers are served, each by 10--12\%, so the AI drivers turn A into a win-win-win, under N and L alike. Drivers who act for themselves leave the zones where A's few offers would leave complying drivers idle, and A does not have to pay them for it.

\subsection{Same driver behavior, different outcomes under R and A}
\label{sec:res-RA}

Having seen the effect of disclosure under each policy on its own, we now put the two policies side by side. In the city, the drivers respond to disclosure in the same way under both policies: disclosure acts through the rebalancing offers, and under L the drivers behave almost identically under the two policies. However, as in the small model (Figure~\ref{fig:movescurve}), the same behaviour leads to different outcomes: from N to L, the four key metrics in Table~\ref{tab:citychange} change in opposite directions under R and A. Under R, driver income, own moves, the service rate and platform profit all fall. Under A, all four rise (Table~\ref{tab:cityallai}). The city thus shows both halves of the argument: disclosure changes the drivers' behaviour, and the same behaviour has different outcomes under different policies. Since the drivers behave alike, the difference lies in the policy.

Under L, the two policies earn almost the same commission at about the same service rate, but A pays far less compensation. A pays both less frequently and less each time: it sends about half as many offers as R, and each accepted offer costs about half as much (Table~\ref{tab:cityallai}). With the same commission and a much smaller compensation expenses, A therefore makes more profit than R. This follows from A's design: it sends an offer only where a moved driver is forecast to bring more commission than it costs, so drivers following its offers cost A almost nothing.

From full compliance to N and then to L, A does steadily better against R. Table~\ref{tab:cityrvsa} compares the two policies three times: with every driver complying, with AI drivers who do not know the rules (N) and with AI drivers who do (L). Across these three columns, A's lead in platform profit grows from 0.3\% to 2.0\% and then 6.1\%, R's lead in service shrinks from 6.4 to 1.4 and then 1.0 points, and R's lead in driver income shrinks from 12.0\% to 3.9\% and then 2.6\%. The reason should be attributed to the R's extra compensation. With complying drivers, each extra USD that R pays over A comes back as 0.99 USD of extra commission, so the two policies tie on profit; under N it brings back 0.51 USD, and under L almost nothing (0.01 USD). The value of anticipation therefore rises as drivers act for themselves, and rises further once they know the rules. Seen from R's side, the same three columns explain why A pulls ahead: as Section~\ref{sec:res-R} shows, R does well only while drivers decline its offers, and disclosure removes the subsidy that the drivers' subsequent free moves gave it.

\begin{table}[htbp]
\centering
\footnotesize
\caption{Difference between the two policies (A minus R) in the city.}
\label{tab:cityrvsa}
\begin{tabular}{l ccc}
\toprule
A minus R & All comply & AI, N & AI, L \\
\midrule
Platform profit ($\Pi$, \%) & $+$0.3 & $+$2.0 & $+$6.1 \\
Service rate (points) & $-$6.4 & $-$1.4 & $-$1.0 \\
Driver income ($I_v$, \%) & $-$12.0 & $-$3.9 & $-$2.6 \\
\midrule
R over A: extra commission per USD of extra compensation & 0.99 & 0.51 & 0.01 \\
\bottomrule
\end{tabular}
\end{table}

Once drivers understand the rules, they take the money the rules offer, and they also take almost every passenger request. Without the rules the agents are selective: under R the offers they accept pay 7.6 USD each on average, against 6.2 USD for a complying driver, and under A 3.5 against 3.2 USD. With the rules they accept almost every offer, at 5.9 and 3.0 USD each. The selective acceptance that surge pricing invites for trips \citep{garg2022surge} thus shrinks for offers under disclosure, and R's loss comes from drivers doing what its rule pays for. In the evening hours (Figure~\ref{fig:cityfree}), the drivers also move more in step with each other under L, under both policies; Appendix~\ref{app:herding} shows that this does not hurt A and asks whether it adds to R's loss.

Thus, the city experiments show that under compulsory disclosure, the design of the rebalancing policy decides the outcome. Disclosure changes what the drivers do in the same way under both policies: they take the platform's offers. R, which pays for a move to every zone that lost requests, loses profit; A, which pays only for moves forecast to bring more than they cost, loses nothing, and its lead over R grows. With self-interested drivers, A becomes a win-win-win, and disclosure amplifies it: platform profit rises, more passengers are served and drivers earn more. From N to L, the driver understands the system better, so the design of the policies matters even more. With a dedicated design, disclosure can guide the drivers to behave in a direction which maximizes the benefits of all parties. Traditional policies which work well under compliant simulations, or self-interested but un-informed drivers scenarios are likely to perform poorly under disclosure, and the design of the policy should be re-evaluated to ensure that it is robust to disclosure.

\subsection{Mixed markets with both AI and compliant drivers}
\label{sec:res-mixed}

With the understanding of the market operates entirely with AI drivers, we next study the scenarios while only part of the drivers use AI agents to make decisions for them. In the mixed scenarios, we tested the city experiments in four cases, each in information condition N and L, with 100 AI drivers with 400 compliant drivers, or 250 AI drivers with 250 compliant drivers. Against the compliant benchmark, every mixed market serves more passengers and raises the platform's profit (Table~\ref{tab:citymixedcells}). In all four test cases, both policies improve in service rate and platform profits monotonically with the share of AI drivers, from 100 to 250 and on to all 500 (Tables~\ref{tab:citymixedcells} and~\ref{tab:cityallai}).

\begin{table}[htbp]
\centering
\caption{City results with 100 or 250 of the 500 drivers as Jev agents and the others complying. $\Pi$ and compensation are in thousand USD a day, driver income per driver over the day. Brackets give the change against compliant benchmark.}
\label{tab:citymixed}
\begin{subtable}{\linewidth}
\centering
\scriptsize
\setlength{\tabcolsep}{3pt}
\caption{Results per cell.}
\label{tab:citymixedcells}
\begin{tabular}{l c c c c c c}
\toprule
 & & & & \multicolumn{3}{c}{Driver income ($I_v$, USD)} \\
\cmidrule(lr){5-7}
Cell & \makecell{Platform\\profit ($\Pi$)} & \makecell{Service\\rate (\%)} & \makecell{Compensation\\paid} & All drivers & AI drivers & \makecell{Compliant\\drivers} \\
\midrule
\rlap{\textit{100 AI drivers, 400 compliant}} \\
R $\times$ N & 138.77 ($+$2.27) & 58.4 ($+$0.5) & 14.02 ($-$1.41) & 591.5 ($+$0.5) & 615.6 ($+$20.8) & 585.4 ($-$4.5) \\
R $\times$ L & 137.67 ($+$1.17) & 58.4 ($+$0.4) & 14.58 ($-$0.85) & 590.2 ($-$0.8) & 601.9 ($+$7.1) & 587.2 ($-$2.7) \\
A $\times$ N & 141.75 ($+$4.89) & 53.7 ($+$2.1) & 2.99 ($-$0.18) & 538.8 ($+$19.0) & 612.2 ($+$87.1) & 520.4 ($+$2.0) \\
A $\times$ L & 141.93 ($+$5.07) & 53.8 ($+$2.2) & 3.08 ($-$0.09) & 540.0 ($+$20.3) & 615.6 ($+$90.4) & 521.1 ($+$2.7) \\
\midrule
\rlap{\textit{250 AI drivers, 250 compliant}} \\
R $\times$ N & 143.19 ($+$6.69) & 58.8 ($+$0.8) & 11.12 ($-$4.30) & 592.3 ($+$1.3) & 601.0 ($+$8.1) & 583.5 ($-$5.5) \\
R $\times$ L & 140.66 ($+$4.16) & 58.6 ($+$0.6) & 12.65 ($-$2.77) & 590.6 ($-$0.3) & 598.8 ($+$5.9) & 582.5 ($-$6.5) \\
A $\times$ N & 145.43 ($+$10.29) & 55.2 ($+$4.2) & 2.83 ($-$0.41) & 553.0 ($+$40.1) & 594.9 ($+$70.6) & 511.1 ($+$9.6) \\
A $\times$ L & 144.60 ($+$9.47) & 55.3 ($+$4.3) & 3.04 ($-$0.20) & 550.4 ($+$37.5) & 597.2 ($+$72.9) & 503.7 ($+$2.2) \\
\bottomrule
\end{tabular}
\end{subtable}

\vspace{8pt}
\begin{subtable}{\linewidth}
\centering
\footnotesize
\caption{Change from N to L (L minus N).}
\label{tab:citymixedchange}
\begin{tabular}{l cccc}
\toprule
 & \multicolumn{2}{c}{100 AI drivers} & \multicolumn{2}{c}{250 AI drivers} \\
\cmidrule(lr){2-3}\cmidrule(lr){4-5}
Change from N to L & R & A & R & A \\
\midrule
Platform profit ($\Pi$, k USD) & \textbf{\boldmath$-$1.10} & $+$0.18 & \textbf{\boldmath$-$2.53} & \textbf{\boldmath$-$0.82} \\
Service rate (points) & 0.0 & $+$0.1 & $-$0.2 & $+$0.1 \\
Commission (k USD) & $-$0.54 & $+$0.27 & $-$1.00 & $-$0.61 \\
Compensation paid (k USD) & $+$0.56 & $+$0.09 & $+$1.53 & $+$0.21 \\
Driver income ($I_v$, USD), all drivers & $-$1.3 & $+$1.2 & $-$1.7 & $-$2.6 \\
\quad AI drivers & $-$13.7 & $+$3.4 & $-$2.3 & $+$2.3 \\
\quad compliant drivers & $+$1.8 & $+$0.7 & $-$1.0 & $-$7.4 \\
\bottomrule
\end{tabular}
\end{subtable}
\end{table}

\begin{figure}[htbp]
\centering
\includegraphics[width=0.8\linewidth]{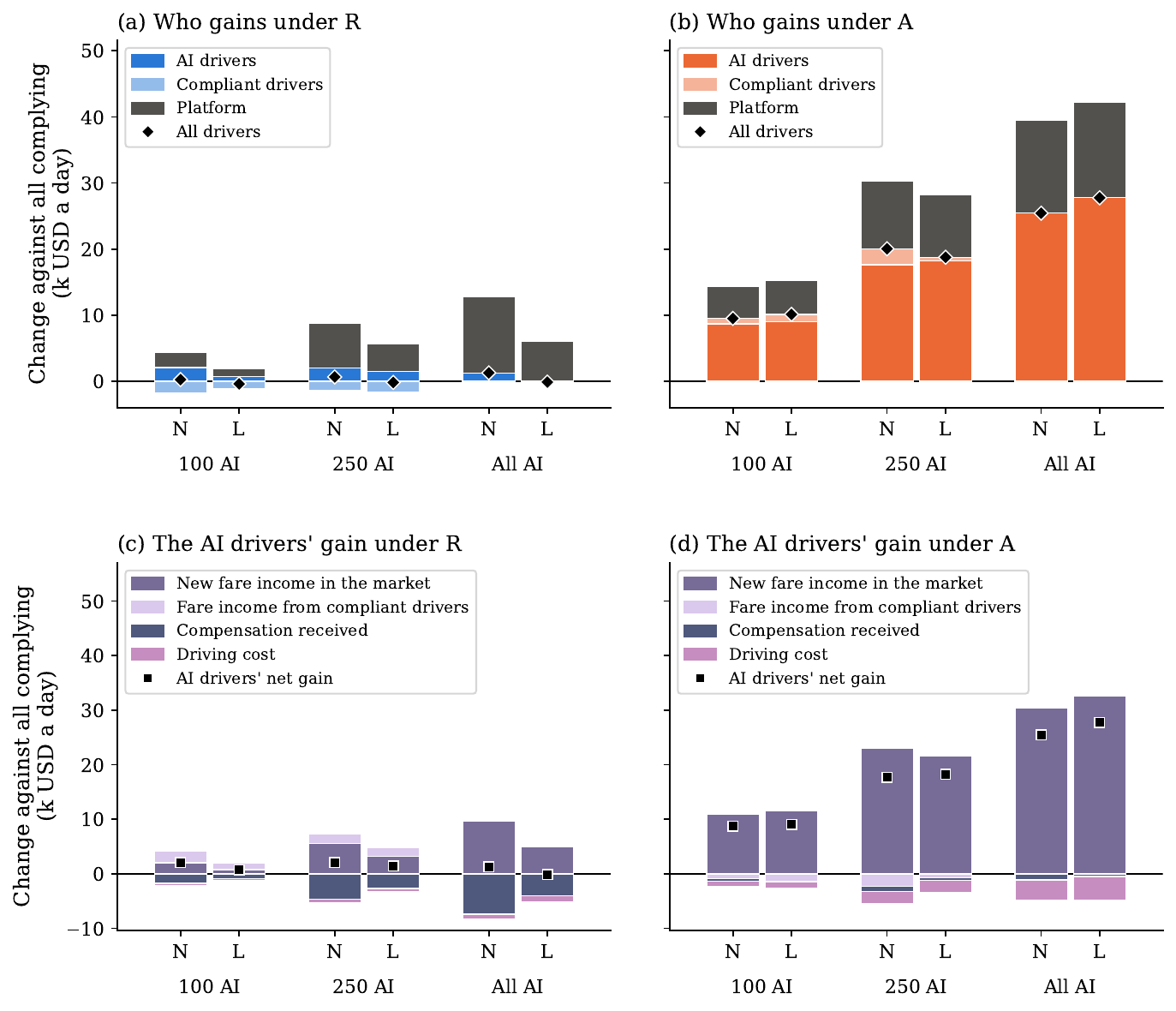}
\caption{Change over the day against everyone complying, by party (a, b) and, for the AI drivers, by source of their gain (c, d). Gains are stacked above zero and losses below it; each row shares one scale.}
\label{fig:citymixed}
\end{figure}

To see how AI drivers and compliant drivers contributes to the improvements of the system, Figure~\ref{fig:citymixed}a,b splits the change against the compliant benchmark across three parties: the AI drivers, the compliant drivers and the platform. 

However, gaming among all drivers are closer to zero-sum under R. In both mixed fleets and under both N and L, the gain of the AI drivers comes from the loss of the compliant drivers (Figure~\ref{fig:citymixed}a). The AI drivers gain 0.7k to 2.1k USD a day and the compliant drivers lose 1.1k to 1.8k, which is close to zero-sum. The platform gains in every case and takes all or nearly all of what drivers and platform gain together. Between 62\% and 73\% of its gain is the compensation it saves when the AI drivers decline its offers; the rest is commission on the extra trips (Table~\ref{tab:citymixedcells}). From N to L, the platform's profit falls in both mixed fleets, as it does when every driver is an AI agent, but in a smaller magnitude (Tables~\ref{tab:citymixedchange} and~\ref{tab:citychange}). As the compliant drivers always follow the platform's instructions, the free moves that subsidize R come only from the AI drivers. With information being disclosed, R pays more compensation for moves it used to get for free, and earns less commission. R's loss thus grows in proportion to the number of AI drivers: 1.1k, 2.5k and 5.4k USD a day with 100, 250 and 500 (Tables~\ref{tab:citymixedchange} and~\ref{tab:citychange}).

Different from R, no group of drivers loses in A, so the game among the drivers is positive-sum. The AI drivers gain together with the platform (Figure~\ref{fig:citymixed}b), with the AI drivers gain more than the platform does. On the common scale of Figure~\ref{fig:citymixed}a,b, the gains under A are much larger than under R. Drivers and platform together gain 14.4k to 30.4k USD a day under A, against 0.8k to 7.4k under R, and the platform gains more under A in every fleet (Table~\ref{tab:citymixedcells}). 

We next examine where the gain of the AI drivers comes from by splitting it into its sources (Figure~\ref{fig:citymixed}c,d). Under R, the AI drivers' extra fare income is largely offset by the compensation they forgo by declining R's offers, and by the cost of their extra driving (Figure~\ref{fig:citymixed}c). What is left roughly equals the fare income they take from the compliant drivers, which again confirms the zero-sum game among the two driver groups. Under A, all of their extra fare income comes from serving additional requests, while the compliant drivers earn more fare income as well (Figure~\ref{fig:citymixed}d). Declined offers cost them little, as A sends fewer rebalancing offers than R and pays less for each (Table~\ref{tab:citybaseline}). As the market serves far more additional trips under A (Table~\ref{tab:citymixedcells}), competition among drivers under A is positive-sum, which reinforces the win-win-win stated in Section~\ref{sec:res-A}: in every mixed fleet, the drivers, the platform and the passengers all gain. The magnitude of gains of AI drivers are also much larger in A, 71 to 90 USD per driver, compare to R, which is only 6 to 21 USD a day (Table~\ref{tab:citymixedcells}). 

\begin{figure}[htbp]
\centering
\includegraphics[width=\linewidth]{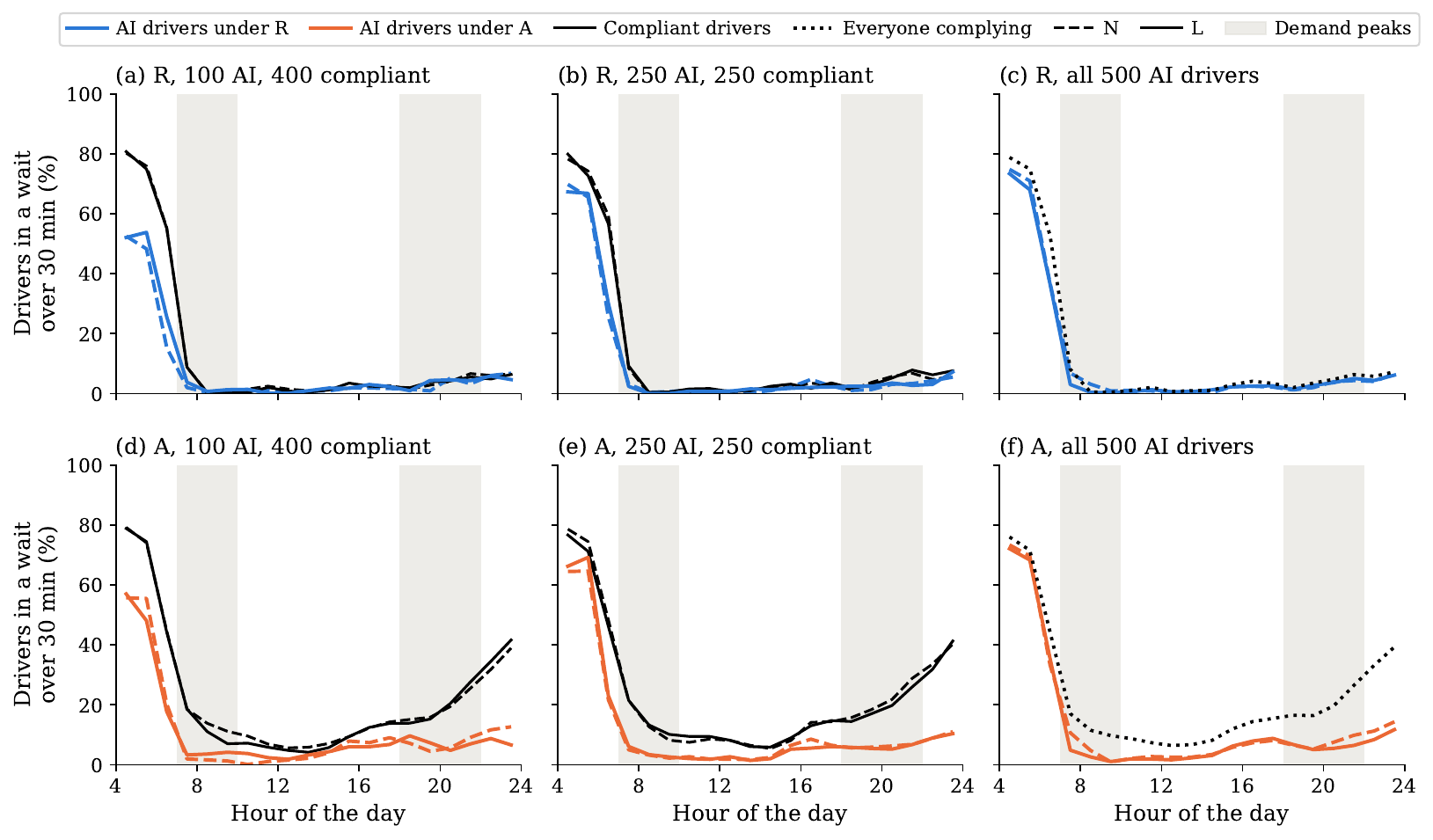}
\caption{Drivers in a wait longer than 30 minutes by hour of the day.}
\label{fig:citylongwaits}
\end{figure}

Another key behavior of AI drivers is their movements on their own. To see how those moves are related with the gains of AI drivers, we first look at how long drivers wait between trips. The gap between finishing the last trip to accepting the next one is defined as a wait, and a long wait is one of over 30 minutes, which are most common when requests are few (Figure~\ref{fig:citylongwaits}). Between 4:00 and 5:00, the hour with the fewest requests (Figure~\ref{fig:cityfree}f), 77\% to 81\% of the compliant drivers are waiting for request. The colored curves reflects the waiting time of AI drivers, which are always less since they move on their own to busier zones, while a compliant driver might stays in quite zones until it is assigned a request or a rebalancing offer. From 100 to 250 and then to all 500 AI drivers, the share of AI drivers in a long wait rises, from 45\% to 50\% and then to 60\% under R, and from 55\% to 60\% and then to 70\% under A. There is a clear diminishing return to having more AI drivers: the more drivers use AI agent, the less each can find a better zone to move to.

Later of the day, different waiting patterns emerge from two policies. Under R, from 8:00 onwards at most 8\% of drivers are in a long wait, and AI and compliant drivers differ little (Figure~\ref{fig:citylongwaits}a--c). Since R sends more offers than A (Table~\ref{tab:citybaseline}), and most of all late in the evening (Figure~\ref{fig:cityfree}a), even compliant drivers can be rebalanced to busier zones, which results a flat tail of waiting time. Under A, long waits among compliant drivers are more common through the day (Figure~\ref{fig:citylongwaits}d--f). Their share climbs from about 15\% at 18:00 to about 40\% in the last hour, although requests remain high (Figure~\ref{fig:cityfree}f). A's offers stay few in these hours (Figure~\ref{fig:cityfree}a), so compliant drivers have little way out. This is where the AI drivers' own moves make a difference: they move on their own to busier zones, and their share in a long wait stays below 10\% through the evening. To better understand the value of these own moves, we next measure what they earn from those unpaid movements.

We first label each zone, in each clock hour, by how quickly an idle driver there finds a trip. The labels come from the benchmark run in which all drivers comply under the same policy to exclude the influence from AI driver's decisions. For every idle decision made in the zone during that hour, we check whether the driver accepted a trip within the next 15 minutes. The zone is poor in that hour if 30\% or fewer of these decisions are followed by a trip, busy if more than 70\% are, and middling otherwise. A busy zone is thus one with plenty of requests for the drivers waiting there and a zone can be poor in one hour and busy in another.

\begin{figure}[htbp]
\centering
\includegraphics[width=0.8\linewidth]{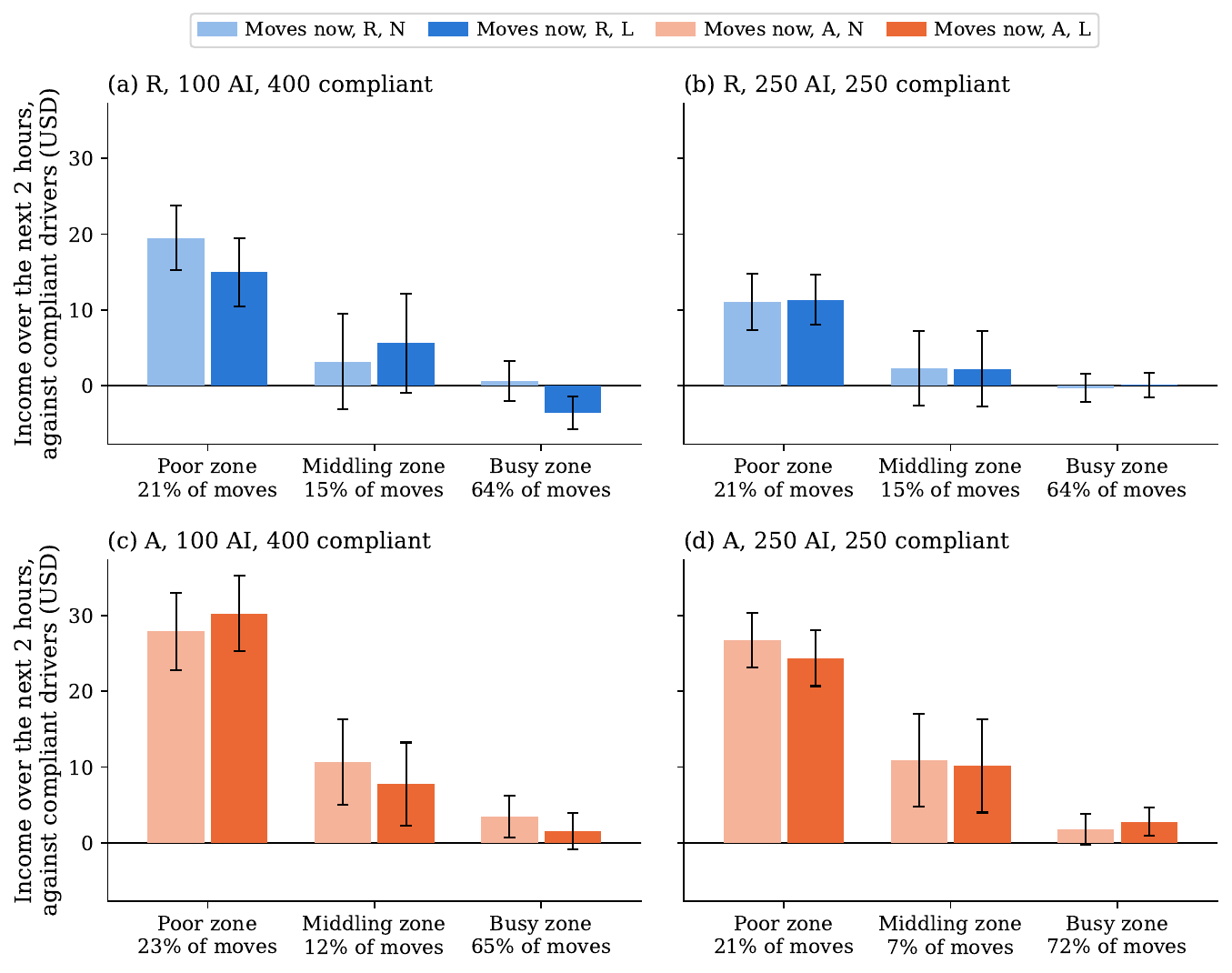}
\caption{Income of AI drivers who move on their own at an idle decision, by the kind of zone they leave, against compliant drivers of the same run in the same zone and hour. Income is the drivers' share of fares and compensation, before driving cost; intervals are 95\% (standard errors clustered by driver).}
\label{fig:citymovevalue}
\end{figure}

\begin{figure}[htbp]
\centering
\includegraphics[width=0.8\linewidth]{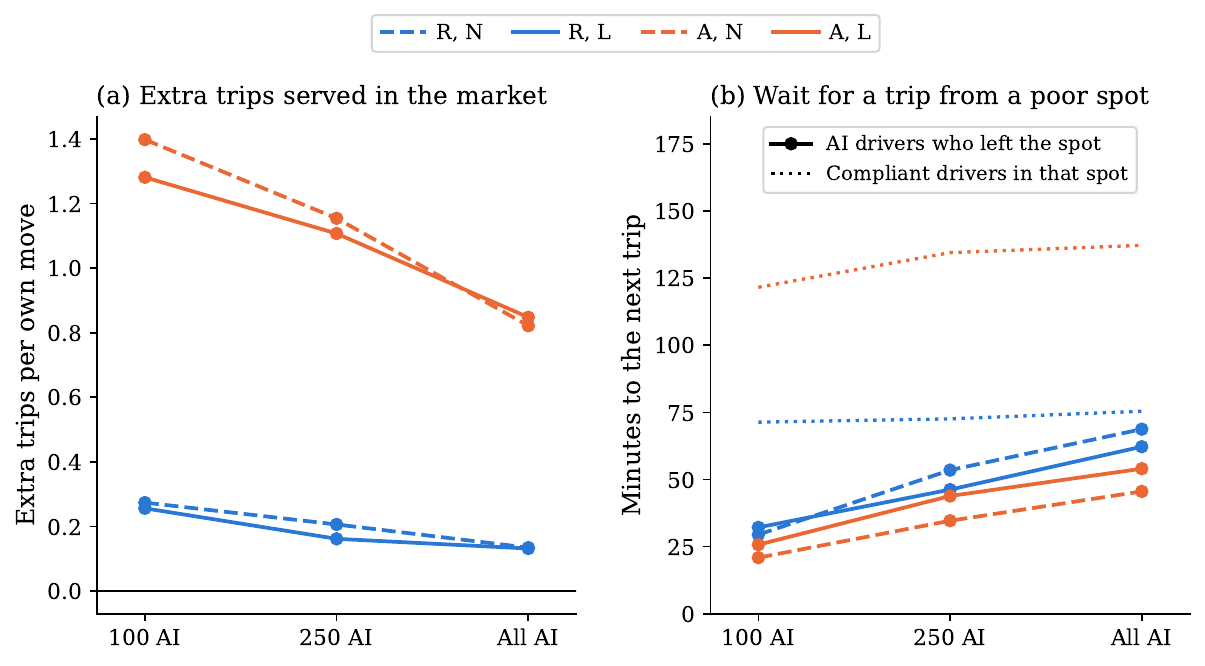}
\caption{Own moves of the AI drivers as their number grows. Extra trips (a) are against everyone complying; in (b), poor zones are those of Figure~\ref{fig:citymovevalue}, and the compliant drivers are those in the same zones and hours when everyone complies, averaged over N and L.}
\label{fig:citymovers}
\end{figure}

Figure~\ref{fig:citymovevalue} gives the income of AI drivers over the next 2 hours, with the benchmark as other compliant drivers in the same zone at the same time. An AI driver who moves on its own from a poor zone earns 13 to 15 USD more under R and about 27 USD more under A. From a middling zone it earns 9 to 11 USD more under A and 3 to 4 USD more under R. From a busy zone it earns 2 to 3 USD more under A, while under R the gain disappears ($+0.1$ USD under N and $-1.7$ under L). The busier the zone, the less an own move gains, under both policies and both conditions. The percentage below each zone label is the total share of movements of AI drivers during the whole simulation, aggregated by the type of zone when such movement happened. Most own moves, 64\% to 72\%, start from busy zones, where a driver is likely to find a trip soon even without moving, so they add little. The other 28\% to 36\% start from poor and middling zones. However, the fewer moves account for most of the income advantage in Figure~\ref{fig:citymovevalue}. Weighting the value of each group by its share of the moves, they make up 75\% to 89\% of the total under A and over 90\% under R. They are also the moves that keep AI drivers out of the long waits of Figure~\ref{fig:citylongwaits}: a compliant driver stays in a poor zone until the platform assigns it a request or a rebalancing offer, while an AI driver leaves and reaches its next trip sooner (Figure~\ref{fig:citymovers}b). Therefore, the value of own moves mainly comes from where they originate, and AI drivers utilize them to turn time they would otherwise spend waiting in poor zones into extra income.

Whether these gains are new to the market depends on the policy (Figure~\ref{fig:citymovers}a). Under R, the whole market serves 0.13 to 0.27 extra trips per own move. The service rate barely rises (Table~\ref{tab:citymixedcells}), and what the AI drivers gain is largely what the compliant drivers lose, which is the zero-sum outcome of Figure~\ref{fig:citymixed}a. Under A, the market serves 0.82 to 1.40 extra trips per own move, at least five times as many as under R. Most of the gain is therefore new to the market, which is the positive-sum outcome of Figure~\ref{fig:citymixed}b.

Under both policies, the benefits to AI drivers diminish as the share of drivers using AI agents increases. From 100 AI drivers to all 500, the extra trips per own move fall from 0.26--0.27 to 0.13 under R and from 1.28--1.40 to 0.82--0.85 under A (Figure~\ref{fig:citymovers}a). While the total number of extra trips grows with the share of AI drivers, each AI driver's gain shrinks from 87--90 USD a day to 51--56 under A, and from 7--21 to at most 3 under R (Tables~\ref{tab:citymixedcells} and~\ref{tab:cityallai}). The wait after a move shows the same pattern (Figure~\ref{fig:citymovers}b). An AI driver who leaves a poor zone always reaches its next trip sooner than compliant drivers in the same zones and hours when everyone complies. However, the saving shrinks from about 41 to 10 minutes under R and from about 98 to 87 under A. This resembles a competitive market: the more drivers pursue the same unserved requests, the less each one can gain by moving.

With a mixture of drivers being compliant and using AI, the two rebalancing policies again respond differently to the information disclosure. Under R, the AI drivers' gain is taken from the compliant drivers, and disclosure costs the platform more as more drivers learn the policies. Thus, the outcome among the drivers is close to a zero-sum game. Under A, the gain is new to the market and shared with the platform and the passengers, and disclosure leaves the platform's profit about unchanged. In contrast, a positive-sum game arises among the drivers, and the same `win-win-win' among the drivers, the platform and the passengers is observed in the mixed fleets as well. Under both policies, own moves pay most for drivers who leave poor zones, and each move brings less as more drivers use AI agents.

\FloatBarrier
\section{Conclusion}
\label{sec:discussion}

The EU Platform Work Directive will oblige ride-hailing platforms to disclose, in plain language, the rules by which their algorithms manage drivers. Plain-language rules are exactly what an LLM can read, and drivers will therefore be able to hand the disclosed rules to an AI assistant that acts in their interest. This study examines two main streams of rebalancing policies, reactive and anticipatory, under such information disclosure with self-interested drivers who are making decisions through LLMs. Experiments compares the behaviors of two LLMs, Jev and Claude Haiku, under a small three-zone model with ten drivers, which can be solved to optimality. Then, this study also test the AI drivers in a more realistic city-scale simulation in Manhattan with 500 drivers. Moreover, to study the competition between AI drivers and compliant drivers, mixed experiments with either 100 or 250 AI drivers out of the same 500 drivers fleet, with the rest as compliant drivers, were conducted in Manhattan. The results show that the information disclosure imposes universal patterns of influences on the behavior of AI drivers, while the same behavior results different outcomes under different policies. 

Comprehension tests show that the LLMs understood the disclosed policies and acted on them consistently. They answered every question correctly, and once the policies were disclosed they followed the platform more closely. In the small model they moved less on their own and earned more, with two unrelated LLMs responding in the same direction; in the city they accepted about 75\% of the platform's rebalancing offers and behaved almost identically under the two policies.

However, the same behaviour had opposite consequences under the two policies. Under the anticipatory rebalancing policy (A), disclosure left drivers, the platform and passengers better off together, which creates a `win-win-win': in the city, the self-interested drivers made every party better off than full compliance would. Under the reactive policy, the effect on the platform depended on how the drivers had behaved before, and in the city disclosure cost the platform 3.7\% of its profit. The reactive rebalancing policy (R) had relied on drivers declining its offers and then moving where it wanted for free; once they understood the rebalancing policies, they accepted the offers, and the platform paid for repositioning it had been getting at no cost, without earning more commission. The anticipatory policy pays only for moves forecast to bring more than they cost, so drivers following it cost it almost nothing, and its lead in profit over the reactive policy grew from 2\% to 6\%.

The above contrast still holds when only some of the drivers use an AI agent to make decisions. With 100 or 250 of the 500 drivers as AI agents, every mixed market served more passengers and earned the platform more than full compliance, and both gains grew with the share of AI drivers. Under the reactive policy, the AI drivers gained roughly what the compliant drivers lost, so the game among the drivers was close to zero-sum. The cost of disclosure to the platform grew with the number of AI drivers: only AI drivers made free moves, and when the policies were disclosed, they took the rebalancing offers instead, which results in the platform losing the free moves it had been getting. Under the anticipatory policy, the AI drivers gained by serving additional requests, the compliant drivers gained slightly too, and the win-win-win held in every mixed fleet.

Own moves play a central role in these gains. Moving on their own let the AI drivers leave poor zones and reach their next trip sooner than compliant drivers, who waited there until the platform assigned them a request or a rebalancing offer. The value of an own move therefore depended mainly on where it started. Moves from poor and middling zones were only 28\% to 36\% of all own moves, yet they accounted for 75\% to 89\% of the AI drivers' income advantage under the anticipatory policy and over 90\% under the reactive policy. Whether the gain was new to the market depended on the policy: each own move brought 0.82 to 1.40 extra trips under the anticipatory policy, against 0.13 to 0.27 under the reactive policy.

The advantage of using an AI agent also diminishes as more drivers used one. Although the market's gains grew with the share of AI drivers, each AI driver's gain fell from 87--90 USD a day with 100 AI drivers to 51--56 with all 500 under the anticipatory policy, and from 7--21 to at most 3 under the reactive policy. As in a competitive market, the more drivers pursued the same unserved requests, the less each one gained. The policy also decided what this spread meant for drivers without an AI agent: under the reactive policy they lost income to the AI drivers, while under the anticipatory policy they earned more as well.

Therefore, with the different policies, the same behaviour can have vastly different outcomes under compulsory disclosure. A policy that performs well with complying drivers, or with self-interested drivers who do not know its policies, can perform poorly once such information is public, and a policy that works only because drivers ignore or misread it will not survive disclosure. Operators should design and evaluate their policies for drivers who understand them and act in their own interest, and an LLM agent reading the disclosed text offers a practical way to do so. For regulators, the results show that law-level disclosure is enough to change what drivers do, and that they do not have to trade transparency against operations: with a well-designed policy, disclosure guides drivers in a direction that benefits every party.

This study leaves several questions open. First, passenger requests were fixed in advance in all experiments, so the results do not show how passengers respond to disclosure. If passengers received the same transparency, they could change their behavior as well, for example how high a fare they are willing to pay, and the outcomes of both policies could shift. LLM agents could represent such passengers, but their behavior would first need to be validated against real-world data. Second, the mixed markets in this study combined compliant drivers only with LLM agents. Adding drivers whose decisions come from RL or from simple heuristics would show whether the contrast between the two policies holds when drivers compete with different strategies. The most direct next step, however, is to design rebalancing policies that are incentive compatible under disclosure, so that the moves informed, self-interested drivers choose are exactly the moves the platform wants and the `win-win-win' becomes a stable outcome. 

\newpage
\bibliographystyle{elsarticle-harv}
\bibliography{references}

\newpage
\appendix

\section{Prompts}
\label{app:prompts}

Each decision of an AI driver sends one prompt, followed by the questions the driver answers. The prompt opens with \texttt{[ROLE]} and \texttt{[PLATFORM RULES]}; under N the rules section is empty, and under A only its rebalancing paragraph differs from R. The prompts call rebalancing offers ``relocation offers''. Jev receives the prompt as one text and answers all questions of a decision in one call, and the driver takes Jev's top choice. Claude Haiku 5.5, used only in the small model, receives \texttt{[ROLE]} and \texttt{[PLATFORM RULES]} as the system message and the rest as the user message, with one call per question. The options of every question are shuffled at each decision. Besides its action, the agent is asked at an idle choice and when holding an offer to predict whether the fare in its zone, or the compensation of its offer, will rise, stay about the same or fall; Haiku is not asked this question.

The full prompts, recorded under R $\times$ L with a stand-in driver that answers like a complying driver, are in the ancillary files of this arXiv version: for the city, \texttt{city\_prompt\_idle\_decision.txt}, \texttt{city\_prompt\_relocation\_offer.txt} and \texttt{city\_prompt\_passenger\_request.txt}; for the small model, \texttt{small\_prompt\_decision\_point.txt} and \texttt{small\_prompt\_passenger\_request.txt}. Their numbers are not from the AI runs; the numbers change at every decision, the wording does not. Below, the city prompt at an idle choice is shown with lists of one value per zone shortened to [\ldots] and only a few of the options for driving to another zone; notes in roman italics are ours.

\subsection{City model}

The rules under L are the five paragraphs of Appendix~\ref{app:rules} followed by the rebalancing paragraph of Figure~\ref{fig:relocation}. At an idle choice, every 15 minutes while the driver waits in a zone, the prompt and its questions are:

\begin{promptbox}[{City model: idle decision}]
[ROLE]\newline
You are an independent ride-hailing driver. Your goal is to maximize your own net income.\newline
[PLATFORM RULES]\newline
\pnote{the rules; empty under N}\newline
[TIME]\newline
clock 08:06:30; the shift ends at 24:00:00\newline
[YOU]\newline
zone 10, available; you keep 0.7 of each fare\newline
your recent history by price interval (oldest first):\newline
07:35-07:40: zone 12, income -0.52, action ACCEPT:9\newline
07:40-07:45: zone 9, income -0.59, action RELOCATING\_WITH\_OFFER\newline
07:45-07:50: zone 9, income 3.34, action RELOCATING\_WITH\_OFFER\newline
07:50-07:55: zone 7, income -0.6, action STAY, request responses [ACCEPT]\newline
07:55-08:00: zone 7, income -0.58, action ON\_TRIP\newline
08:00-08:05: zone 10, income -0.8, action ON\_TRIP\newline
[MARKET]\newline
your current zone and city totals, recent history by price interval (oldest first):\newline
07:35-07:40: zone 10, requests 10, online idle 1.5, fare 5.28, offer compensation by destination [\ldots], matched requests 9, city requests 250, city online idle 33.7, city matched requests 132\newline
\pnote{four more price intervals}\newline
08:00-08:05: zone 10, requests 14, online idle 0.6, fare 4.32, offer compensation by destination [\ldots], matched requests 14, city requests 215, city online idle 57.3, city matched requests 143\newline
[DEMAND]\newline
current observation, last hour (07:05-08:05), by zone: average fare per km [\ldots]; requests [\ldots]\newline
history experience, next hour (08:05-09:05), by zone: expected requests [\ldots], the average at this time of day on earlier weekdays (Mon-Thu); today's requests vary around it.\newline
distance from you, by zone: road km to the zone's centre [\ldots]\newline
your current location is zone 10
\tcblower
action: Choose your action now. MOVE\_Zxx means driving on your own to the centre of zone xx: no compensation, and you stay available for requests and offers along the way. Zone columns: fare = current fare per km of a trip starting there; offer = what a relocation offer from where you are to that zone would pay you (paid only to the holder of that offer); idle = drivers available there, waiting or passing through; requests = requests since the last price update; match = recent share of idle drivers who got a request; travel = time in seconds, road distance and driving cost from where you are to the zone's centre, where a move there ends.\newline
\hspace*{1em}STAY: Stay in your zone, available for requests. zone 10 | fare 5.98 | idle 3 | requests 14 | match 1\newline
\hspace*{1em}MOVE\_Z09: zone 9 | fare 6.31 | offer 3.34 | idle 1 | requests 32 | match 1 | travel 570 s, 2.11 km, cost 0.49\newline
\hspace*{1em}MOVE\_Z12: zone 12 | fare 4.32 | offer 1.34 | idle 14 | requests 38 | match 1 | travel 591 s, 1.24 km, cost 0.28\newline
\hspace*{1em}\pnote{one MOVE option for each of the other 26 zones}
\par\medskip
prediction: Predict the fare in your current zone at the next price update, compared with its current value.\newline
\hspace*{1em}DOWN: Decrease\newline
\hspace*{1em}UP: Increase\newline
\hspace*{1em}FLAT: Stay approximately the same
\end{promptbox}

When the driver is sent a rebalancing offer, the prompt adds \texttt{[YOUR OFFER]} between \texttt{[YOU]} and \texttt{[MARKET]}, with the origin, the destination zone's centre and the compensation. The action question then offers to accept the offer, to decline it and stay, or to decline it and drive to one of the other zones, and the prediction question asks about the compensation of the offered route. At a passenger request, \texttt{[YOUR REQUEST]} takes its place, with the pickup and drop-off zones, the trip's distance and time, the pickup time, the fare and the driver's share, and the number of earlier rejections; the options are to accept the request, to reject it and stay, or to reject it and drive to one of the other zones, and there is no prediction question.

\subsection{Small model}

The small model's prompt has the same structure. Time runs in periods, \texttt{[MARKET]} covers all three zones, and \texttt{[DEMAND]} gives the requests of each completed period by origin and destination. Two sections are added: \texttt{[QUEUE]}, the driver's place in its zone's queue, and \texttt{[EXPECTED DEMAND]}, the mean requests of the demand pattern for this and the next two periods. The prompt says nothing about the other drivers. The rules under L are:

\begin{ruleboxlong}[{[PLATFORM RULES] under L in the small model: paragraphs 1--5, identical for R and A, then the rebalancing paragraph of R or A}]
How each period works. The three zones lie on a line. A passenger trip or a move to a neighbouring zone takes one period; between the two end zones it takes two periods, passing through the middle zone without stopping. Drivers who are still travelling make no decisions and get no requests until they arrive. At the start of each period, the platform sets fares and relocation compensation, plans relocation and sends relocation offers. Then all drivers who are not travelling, including those who rejected a request in the previous period, choose at the same time: accept their offer if they hold one, stay, or move to another zone. Then passenger requests are assigned to the drivers who stayed.

\medskip
Queues and passenger requests. In each zone, waiting drivers form a queue, longest wait first; relocation offers and passenger requests go to the front of the queue. Drivers arriving in a zone join the back; a driver who rejects a request goes to the very back. A rejected request goes to the next driver in line, never to anyone who already rejected it; it is lost after a fixed number of rejections or when no driver is left. After rejecting, you get no more requests until the next period. A passenger's fare is fixed when the request is made and paid when the trip is completed.

\medskip
Relocation offers. Each driver gets at most one offer per period, valid only in that period. Accepting locks its compensation, which is paid when you arrive. Only the driver holding an offer receives its compensation; moving on your own earns none. If you do not accept, the offer expires. Offers are planned again in the next period, so you may get a new one, possibly with a different compensation.

\medskip
Fares and compensation. Each period, the platform measures each zone's shortage: the share of the zone's requests in the previous period that the drivers counted for the zone could not cover. Counted drivers are those who stayed in the zone and were still idle after requests were assigned (the ``online idle'' numbers in the market history), plus those committed to arrive there this period, either finishing a passenger trip there or travelling there on an accepted offer. Drivers who rejected a request, and drivers moving there on their own, are not counted. There is no shortage when the counted drivers cover all requests or there were no requests, and the shortage is greatest when no driver is counted. The fare for each period of a trip starting in a zone is a fixed base fare plus an amount that rises with the zone's shortage, so it is lowest when there is no shortage and highest when no driver is counted; a two-period trip pays twice that. An offer's compensation is a fixed base amount for each period of travel plus an amount that rises with the shortage of its destination zone. Nothing else affects fares and compensation. Drivers who stayed idle and committed arrivals count the same, and one counted driver fewer raises a zone's shortage, and so its fare and the compensation of offers into it, at least as much as one request more does.

\medskip
How your actions affect the platform's decisions. The data about you that these decisions use are your zone, your place in the zone's queue, and how you respond to passenger requests and relocation offers. You are counted in a zone for the next period's prices if you stay there and get no passenger request, and for the prices of the period you arrive if you accept a passenger request or a relocation offer that ends there; each counted driver lowers the zone's shortage, and so its fare and the compensation of offers into it, until the counted drivers cover all its requests. If you reject a passenger request, you go to the very back of your zone's queue, get no more requests this period, and are not counted in your zone, so its next fare and the compensation of offers into it can be higher. If you move to another zone on your own, you join the back of its queue and are not counted there in the prices of the period you arrive.

\medskip
\textit{Under R:} Relocation. Each period, the platform counts the requests lost in each zone in the previous period. Starting from the drivers now in each zone, including those who rejected a request in the previous period, it sends drivers to cover these lost requests with as little travel as possible: drivers already in that zone cover it first, without an offer, and the rest come from other zones through relocation offers, one per lost request, while drivers are available. A request that is lost after you reject it counts among your zone's lost requests. The lost requests decide where offers go and how many; travel decides which zones the drivers come from.

\medskip
\textit{Under A:} Relocation. Each period, the platform forecasts the requests in each zone for the next several periods. Starting from the drivers now in each zone and those on their way to one, not counting those who rejected a request in the previous period, it plans driver movements over those periods to maximize its expected commission from fares minus the compensation it pays. The forecasts and the drivers in each zone decide where drivers will be short; comparing the commission a moved driver is expected to bring with the compensation decides whether to move one. Only the movements planned for the current period are sent as relocation offers; the plan is redone every period.
\end{ruleboxlong}

At a passenger request, \texttt{[YOUR REQUEST]} replaces \texttt{[YOUR OFFER]}, \texttt{[MARKET]} first lists the current zone table, and there is no \texttt{[QUEUE]} and no prediction question.

\section{The disclosed rules shared by R and A (city version)}
\label{app:rules}

\begin{ruleboxlong}[{[PLATFORM RULES] under L, paragraphs 1--5, identical for R and A (city)}]
How the platform works over time. The city is divided into 29 zones of different sizes. Travel between zones takes different times, depending on the road distance. The zone table shows the travel time in seconds, the road distance and the driving cost from where you are to the centre of each zone, where a move to that zone ends; a passenger request shows its own pickup and trip times. Passenger requests arrive at any time and are matched to drivers every few seconds. Fares and relocation compensation are updated at regular intervals of a few minutes, and relocation is planned at regular intervals. While you wait in a zone, every 15 minutes (counted from your last choice of any kind) you choose to stay in your zone or drive to another zone. When a passenger request is assigned to you, you choose at once to accept it, or to reject it and then stay in your zone or drive to another zone. When a relocation offer is sent to you, you choose at once to accept it, or to decline it and then stay in your zone or drive to another zone. Drivers carrying a passenger or driving to a zone on an accepted offer make no choices and get no requests or offers until they arrive. If you drive to another zone on your own, you stay available along the way and can be matched to a request or sent a relocation offer in any zone you pass through. You make no 15-minute choice while driving on your own; the drive ends when you arrive or when you accept a request or an offer.

\medskip
Requests and matching. Every few seconds, the platform matches waiting requests to idle drivers. It matches as many requests as it can, each to a driver who can reach the passenger within a few minutes, and prefers the drivers with the shortest pickup time. A rejected request goes to another driver close enough, never to anyone who already rejected it; it is lost after a fixed number of rejections or when no driver is close enough. If you reject a request and stay in your zone, you remain available and can be matched to the next request. The fare is fixed when the request is made and paid when the trip is completed.

\medskip
Relocation offers. When the platform plans relocation, it sends offers to idle drivers in the zones that drivers should leave, at most one per driver. You answer an offer as soon as it reaches you, and it is valid only for that answer. Accepting locks its compensation, which is paid when you arrive at the destination zone's centre. Only the driver holding an offer receives its compensation; moving on your own earns none. If you decline, the offer goes at once to another idle driver in your zone, never back to you, while such drivers are left; when the platform next plans relocation you may get a new offer, possibly with a different compensation.

\medskip
Fares and compensation. At each update, the platform measures each zone's shortage: the share of the zone's requests since the last update that the drivers counted for the zone could not cover. Counted drivers are those available in the zone on average since the last update (the ``online idle'' numbers in the market history), plus those committed to arrive there before the next update, either finishing a passenger trip there or driving there on an accepted offer. A driver moving on their own is counted in whichever zone they are passing through, not in the zone they are heading to. There is no shortage when the counted drivers cover all requests or there were no requests, and the shortage is greatest when no driver is counted. A trip's fare is its road distance times a rate per kilometre set for the zone where it starts: a fixed base rate plus an amount that rises with the zone's shortage. An offer's compensation is your road distance from where you are to the destination zone's centre times a quarter of the destination zone's rate per kilometre, so it rises with the shortage of the destination zone and always pays more than the drive costs. Nothing else affects fares and compensation. Available drivers and committed arrivals count the same, and one counted driver fewer raises a zone's shortage, and so its fare and the compensation of offers into it, at least as much as one request more does.

\medskip
How your actions affect the platform's decisions. The data about you that these decisions use are your zone, your location in it (which decides how quickly you can reach a new passenger), and how you respond to passenger requests and relocation offers. You are counted in a zone while you are available there, and as a committed arrival if you accept a passenger request or a relocation offer that ends there; each counted driver lowers the zone's shortage, and so its fare and the compensation of offers into it, until the counted drivers cover all its requests. If you reject a passenger request and stay, you remain available and are still counted in your zone, so rejecting does not change its fare; the request goes to another driver, and is lost if no driver takes it. If you drive to another zone on your own, you stay available and are counted in each zone you pass through, so you count in your destination zone only once you reach it.
\end{ruleboxlong}

The sixth paragraph, on rebalancing, is the only one that differs between R and A; both versions are shown in Figure~\ref{fig:relocation}. Under N the \texttt{[PLATFORM RULES]} section is empty and the rest of the prompt is unchanged.

In every condition the city prompt also has a \texttt{[DEMAND]} section. For each zone it gives the average fare per km and the number of requests over the last hour, the expected requests over the next hour, and the road distance from the driver to the zone's centre. The expected requests are the training-day averages that policy A's forecast also uses. The section reads:

\begin{rulebox}[{[DEMAND] in the city prompt, every condition; each [\ldots] lists one value per zone}]
current observation, last hour (\ldots), by zone: average fare per km [\ldots]; requests [\ldots]\\
history experience, next hour (\ldots), by zone: expected requests [\ldots], the average at this time of day on earlier weekdays (Mon-Thu); today's requests vary around it.\\
distance from you, by zone: road km to the zone's centre [\ldots]\\
your current location is zone $z$
\end{rulebox}
\section{Parameter values}
\label{app:params}

Table~\ref{tab:params} gives the value of every parameter of the two instances, in the notation of Table~\ref{tab:notation}. 

\begin{table}[H]
\centering
\footnotesize
\caption{Parameter values of the two instances.}
\label{tab:params}
\setlength{\tabcolsep}{3.5pt}\renewcommand{\arraystretch}{0.9}
\begin{tabular}{@{}L{3.75cm} L{1.55cm} L{4.4cm} L{5.55cm}@{}}
\toprule
Parameter & Symbol & Small model & City (Manhattan) \\
\midrule
\multicolumn{4}{@{}l}{\textbf{Network and clock}} \\[1pt]
Zones & & 3 on a line (residential, middle, CBD) & 29 hexagons of 1,000\,m radius on the road network \\
Travel time between zone centres & $t_{od}$ & 1 period between neighbours, 2 between the end zones & along the road network \\
Unit of distance & $\ell_{od}$ & one zone spacing, driven in one period & road km \\
Decision step & $\Delta T$ & one period of 5 min & 15\,s \\
Price interval & $\Delta T_p$ & one period & 5 min \\
Planning interval & $\Delta T_r$ & one period & 10 min (two price intervals) \\
Horizon of A & $H$ & 6 periods (30 min) & 6 price intervals (30 min) \\
Simulated day & & 36 periods after 6 warm-up periods & 4:00--24:00, after a warm-up hour from 3:00 in which every driver complies \\
Idle-decision interval & $\Delta T_i$ & one period & 15 min \\
Own and market history in the prompt & & last 6 periods & last 6 price intervals \\
Expected requests in the prompt & & this and the next 2 periods & the next hour \\
\addlinespace[3pt]
\multicolumn{4}{@{}l}{\textbf{Demand}} \\[1pt]
Mean requests per period & $\lambda_0$, $\lambda_1$ & 3.5 and 1.5 & -- \\
Peak periods; peak width & $t^{\text{am}}$, $t^{\text{pm}}$; $\sigma$ & 9 and 26; 5 periods & -- \\
Origin--destination shares, morning & $w_{od}(t)$ & 0$\to$2 60\%, 1$\to$2 20\%, 0$\to$1 10\%, 2$\to$0 and 2$\to$1 5\% each; reversed in the evening, linear in between & -- \\
Requests & & 179, one frozen Poisson draw & NYC yellow-taxi records, March 2016; about 53,000 between 4:00 and 24:00 \\
Sampling share & $\rho$ & -- & 0.15 \\
Test day; training days & & -- & Wed 23 March 2016; 1--3 and 7--10 March 2016 \\
Forecast used by A & $\hat q_{odk}$ & $\lambda_{od}(t)$ of the coming periods & mean of the training days at the same time of day, times $\rho$ \\
\addlinespace[3pt]
\multicolumn{4}{@{}l}{\textbf{Fleet and matching}} \\[1pt]
Fleet size & $f_V$ & 10 & 500 \\
Vehicle capacity & $c$ & 1 & 1 \\
Maximum pick-up waiting time & $t_w^{\max}$ & -- (only drivers in the request's zone) & 5 min \\
Order of assignment & & front of the zone's queue first & shortest pickup first \\
Rejections before a request is lost & & 3 & 3 \\
\addlinespace[3pt]
\multicolumn{4}{@{}l}{\textbf{Prices, payments and costs}} \\[1pt]
Base fare rate & $p$ & 0.4 per unit of distance & 4.32 USD per km (12 USD for the mean 2.78-km trip, the 2016 NYC taxi fare) \\
Fare range & & 0.4--0.8 per unit of distance & 4.32--8.63 USD per km \\
Commission & $\theta$ & 30\% & 30\% \\
Driving cost & $\beta_\ell$ & 0.05 per unit of distance & 0.23 USD per km \\
Time cost & $\beta_T$ & 0 & 4.64 USD per hour online \\
Compensation, Eq.~\eqref{eq:compgen} & $b_0$, $\kappa_{vd}$ & $b_0=0.1$ per unit of distance; $\kappa_{vd}=0.15$ for every move & $b_0=\alpha\,p$, $\kappa_{vd}=\alpha\,p\,\ell_{vd}$, so $b_{vd}=\alpha\,p\,(1+g_d)\,\ell_{vd}$ \\
Share of the fare rate & $\alpha$ & -- & 0.25 (1.08--2.16 USD per km) \\
Value of waiting time & $\beta_w$ & -- & 4.64 USD per hour \\
\addlinespace[3pt]
\multicolumn{4}{@{}l}{\textbf{Runs}} \\[1pt]
Runs & & 15 per cell (3 sets of start positions $\times$ 5 repeats) & Compliant benchmark: 5 repeats on the test day; each AI run is paired with the repeat that has the same random seeds \\
\bottomrule
\end{tabular}
\end{table}

\section{LLM models: versions and costs}
\label{app:models}

Table~\ref{tab:models} lists, for every set of runs reported in the paper, the model and its cost per 1,000 API calls.

\begin{table}[H]
\centering
\footnotesize
\caption{Models of the runs reported in the paper and their cost. One API call answers one decision of an AI driver. Costs are as recorded by the runs, at 0.042 USD per million input tokens for Jev and 0.10 and 0.50 USD per million input and output tokens for Claude Haiku 5.5.}
\label{tab:models}
\begin{tabular}{L{4.4cm} L{3.0cm} r L{1.6cm}}
\toprule
Experiment & Model & \makecell{Cost per 1,000\\calls (USD)} & Dates \\
\midrule
Small model, all ten drivers AI & Jev 1.13.0, option of highest probability & 0.119 & 17 Oct \\
Small model, all ten drivers AI & Claude Haiku 5.5, effort xhigh & 1.336 & 17 Oct \\
City, all 500 drivers AI & Jev 1.13.0, option of highest probability & 0.228 & 23 Oct \\
\bottomrule
\end{tabular}
\end{table}

\section{Comprehension test: question topics}
\label{app:questions}

Tables~\ref{tab:questions-small} and \ref{tab:questions-city} list the questions as asked, with their options (shuffled at every call; listed here in bank order), the correct answer and, for Jev, the probability it put on the correct option. Haiku returns no probabilities; it answered every question correctly.

{\scriptsize
\setlength{\tabcolsep}{4pt}
\begin{longtable}{@{}r L{7.0cm} L{5.85cm} c c@{}}
\caption{Comprehension questions of the small model (12 per policy), rules text only, disclosure L. The dispatch question has a different correct answer under R and A, marked (R) and (A). Jev 1.13.0 on 12 October 2026: probability on the correct option; Claude Haiku 5.5 (xhigh), asked the same day, answered every question correctly under both policies.}\label{tab:questions-small} \\
\toprule
\# & Question & Options (correct in bold) & Jev R & Jev A \\
\midrule
\endfirsthead
\toprule
\# & Question & Options (correct in bold) & Jev R & Jev A \\
\midrule
\endhead
\bottomrule
\endfoot
1 & In a strictly positive shortage away from payment caps, with all other inputs fixed, fewer online idle vehicles make fare: & \textbf{Increase}; Unchanged; Decrease & 1.00 & 1.00 \\
2 & Hold requests and committed arrivals fixed. In a positive shortage away from caps, a rejection reducing measured online idle supply makes next period fare: & \textbf{Increase}; Unchanged; Decrease & 1.00 & 1.00 \\
3 & In a positive shortage away from caps, increasing the committed supply used for pricing makes fare: & Increase; Unchanged; \textbf{Decrease} & 0.99 & 1.00 \\
4 & In a positive shortage with positive included supply, away from caps, increasing the demand input used in the rule makes fare: & \textbf{Increase}; Unchanged; Decrease & 0.98 & 0.98 \\
5 & Which demand input does the fare rule use? & \textbf{Previous-period observed requests}; Forecast potential demand & 1.00 & 1.00 \\
6 & What decides where relocation offers send drivers? & \textbf{Requests left unserved recently} (R); \textbf{Forecast demand over the coming periods} (A) & 1.00 & 1.00 \\
7 & When is the relocation compensation amount locked? & \textbf{When the driver accepts}; When the driver arrives & 1.00 & 1.00 \\
8 & When is accepted relocation compensation paid? & At acceptance; \textbf{At arrival} & 1.00 & 1.00 \\
9 & Does independent relocation receive compensation? & Yes; \textbf{No} & 1.00 & 1.00 \\
10 & Can a paused driver be assigned another request before the next decision point? & Yes; \textbf{No} & 0.94 & 0.92 \\
11 & Does seeing a route's compensation imply you can claim it without holding its offer? & Yes; \textbf{No} & 1.00 & 1.00 \\
12 & When is a passenger fare settled? & When requested; \textbf{When delivered} & 0.99 & 1.00 \\
\end{longtable}

\begin{longtable}{@{}r L{7.0cm} L{5.85cm} c c@{}}
\caption{Comprehension questions of the city (17 per policy), rules text only, disclosure L. Jev 1.13.0 on 22 October 2026, archived in the frozen package: probability on the correct option.}\label{tab:questions-city} \\
\toprule
\# & Question & Options (correct in bold) & Jev R & Jev A \\
\midrule
\endfirsthead
\toprule
\# & Question & Options (correct in bold) & Jev R & Jev A \\
\midrule
\endhead
\bottomrule
\endfoot
1 & A zone has a shortage. With everything else fixed, if fewer drivers are counted as available in it, its fare per kilometre at the next price update: & \textbf{Increase}; Unchanged; Decrease & 1.00 & 1.00 \\
2 & Hold everything else fixed. A driver rejects a passenger request and stays in the zone. The zone's next fare: & Increase; \textbf{Unchanged}; Decrease & 0.94 & 0.98 \\
3 & A zone has a shortage. With everything else fixed, if more drivers are committed to arrive there before the next price update, its next fare: & Increase; Unchanged; \textbf{Decrease} & 1.00 & 1.00 \\
4 & A zone has a shortage and some counted drivers. With everything else fixed, if more requests start there before the next price update, its next fare: & \textbf{Increase}; Unchanged; Decrease & 0.96 & 0.99 \\
5 & Which demand does the fare rule use? & \textbf{Requests that started in the zone since the last price update}; Forecast requests for the coming minutes & 1.00 & 1.00 \\
6 & What decides where relocation offers send drivers? & \textbf{Requests lost in each zone since the platform's last plan} (R); \textbf{Forecast requests in each zone for the near future} (A) & 1.00 & 1.00 \\
7 & How is a relocation offer's compensation set? & \textbf{Your road distance to the destination zone's centre times a quarter of that zone's rate per kilometre}; Only the driving cost of the trip; A fixed amount for every offer & 1.00 & 1.00 \\
8 & With everything else fixed, if the destination zone's shortage rises, the compensation of offers into it: & \textbf{Increase}; Unchanged; Decrease & 1.00 & 1.00 \\
9 & When is the relocation compensation amount locked? & \textbf{When the driver accepts}; When the driver arrives & 1.00 & 1.00 \\
10 & When is accepted relocation compensation paid? & When the driver accepts; \textbf{When the driver arrives at the destination zone's centre} & 1.00 & 1.00 \\
11 & Does driving to another zone on your own earn compensation? & Yes; \textbf{No} & 1.00 & 1.00 \\
12 & If you decline a relocation offer, what happens to it? & \textbf{It goes at once to another idle driver in your zone, never back to you}; It expires and is offered again at the next price update; It stays open for you until the platform next plans relocation & 1.00 & 1.00 \\
13 & Does seeing a route's compensation imply you can claim it without holding its offer? & Yes; \textbf{No} & 1.00 & 1.00 \\
14 & After rejecting a passenger request and staying in your zone, can you be matched to another request right away? & \textbf{Yes}; No & 1.00 & 1.00 \\
15 & While you drive to another zone on your own, in which zone are you counted for prices? & \textbf{The zone you are passing through}; The zone you are heading to; No zone until you arrive & 1.00 & 1.00 \\
16 & While you wait in a zone, how often do you choose to stay or move? & \textbf{Every 15 minutes, counted from your last choice of any kind}; At every price update; Whenever you want & 1.00 & 1.00 \\
17 & When is a passenger fare paid? & When the request is made; \textbf{When the trip is completed} & 1.00 & 1.00 \\
\end{longtable}

}

\section{City model: herding}
\label{app:herding}

\begin{figure}[htbp]
\centering
\includegraphics[width=0.62\linewidth]{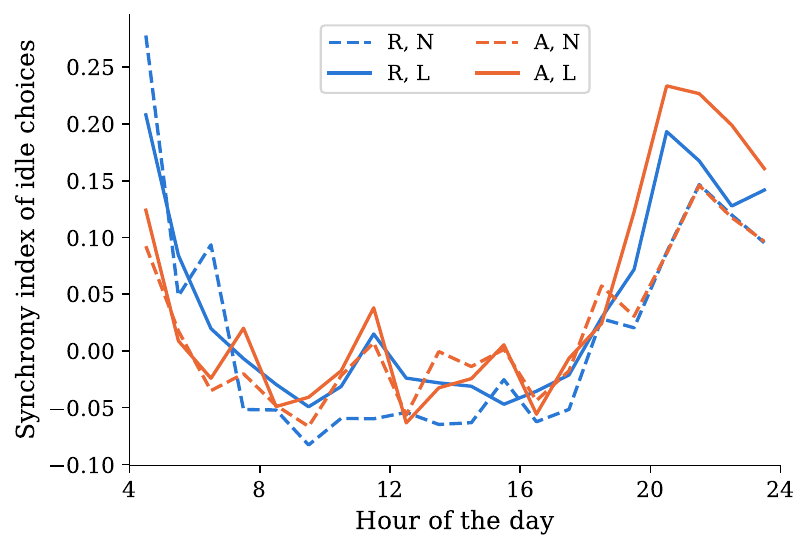}
\caption{Synchrony of the drivers' idle choices in the city by hour of the day: the mean over groups of at least two drivers deciding in the same zone and time window of length $\Delta T_p$. In each group, the index is the largest share of drivers taking the same deviating action, minus its expectation when the drivers draw independently from the run's action distribution (Monte Carlo).}
\label{fig:cityherding}
\end{figure}

All 500 drivers in the city run the same model, so they may act alike for that reason alone: agents that share one model can make the same choice in the same situation and crowd into the same zones \citep{kleinberg2021monoculture}. Disclosure could add to this, since every driver then reads the same rules. This appendix checks whether the drivers move in step, using the synchrony index of their idle choices (Figure~\ref{fig:cityherding}): among drivers in the same zone choosing at the same time, the share that makes the same move, minus the share expected if each chose independently. Over the whole day the index is small, and disclosure raises it by the same amount under both policies, from 0.016 to 0.043 under R and from 0.021 to 0.048 under A (Table~\ref{tab:citychange}). The drivers act alike only at certain hours: the index is high at the start of the day and in the evening and slightly below zero in between, and the rise under L comes from the evening, where at 20:00 it reaches about 0.19 against 0.09 under R and 0.23 against 0.09 under A. Yet A $\times$ L, the cell with the most synchrony, has the highest platform profit of the four: at 500 drivers, moving in step does not turn A's gains into losses.

The hours of high synchrony are not those of low demand: the evening, where the rise under L lies, is the busiest part of the day (Figure~\ref{fig:cityfree}f). They are the hours in which the drivers move most on their own (Figure~\ref{fig:cityfree}d and e), and two reasons combine. First, the index's baseline is the whole day's mix of choices, so hours in which drivers move more than their daily average score high by construction, and the negative values at midday do not mean that drivers avoid each other. Second, agents that run the same model act alike when the situation gives one clear signal of where to go: at 4:00 the fleet starts out of position after the warm-up and heads the same way, under N and L alike, and under R the index is higher then than at any other hour. Both reasons may lie behind the evening rise under L. From 20:00 to 21:00 the drivers make more own moves under L than under N, about 340 against 270 under R and 360 against 250 under A, so part of the rise comes from moving more. The rest may come from the drivers reading the same disclosed rule, that fares rise with the shortage, into the same signal; the figures cannot separate the two.

Whether moving in step harms R is open. The evening is also when R's compensation is wasted (Section~\ref{sec:res-R}). R's plan sends one driver per lost request (Eq.~\eqref{eq:R}) without counting drivers already on their way to that zone, and the shortage that sets the fare counts drivers heading to a zone only if they are committed to arrive there (Eq.~\eqref{eq:shortage}), so neither sees drivers moving there on their own. When informed drivers move in step towards the zones that lost requests, R may pay further drivers to go to the same zones, which then hold more drivers than requests. A sends few offers and is barely exposed. The same rise in synchrony would then have different consequences under the two policies, as disclosure has in the rest of the city results. The figures do not settle this: in the evening under L few own moves follow R's planned pairs of zones (Figure~\ref{fig:cityfree}d), but drivers from other zones may still head to the zones R sends paid drivers to, which comparing the destinations of the evening's own moves with those of R's offers would test. The second measure of herding, the share of own moves beyond the platform's plan, rises from 0.79 to 0.93 under R and from 0.82 to 0.96 under A (Table~\ref{tab:citychange}), for different reasons under the two policies. Under R it rises mostly because the own moves that followed the plan became accepted offers: they fall by about 650 a day, while own moves beyond the plan rise by only about 150. Under A the two parts are about equal: own moves following the plan fall by about 550 a day and those beyond it rise by about 600, most of them in the evening (Figure~\ref{fig:cityfree}d and e).

\end{document}